\documentclass[sigconf]{acmart}

\usepackage{color,soul}

\usepackage{amssymb,amsfonts}
\usepackage{textcomp}
\usepackage{xspace}
\usepackage{multirow}
\usepackage{subfig}
\usepackage[linesnumbered,ruled,vlined]{algorithm2e}
\usepackage{pifont}
         
\newcommand{\name}{\textit{Phi}\xspace}

\newcounter{rlabelno}

\newcommand{\revise}[2]{#2}

\newcommand{\retag}[1]{\hyperref[#1]{\textcolor{red}{\textbf{#1}}}} 

\newcommand{\Fig}[1]{Fig.~\ref{#1}}

\newcommand{\Tbl}[1]{Tbl.~\ref{#1}}
\newcommand{\Sec}[1]{Sec.~\ref{#1}}
\newcommand{\Reb}[1]{\textcolor{red}}
\newcommand{\Alg}[1]{Alg.~\ref{#1}}

\newcommand{\revision}[1]{#1}
\newcommand{\highlight}[1]{#1}

\definecolor{mycolorblue}{RGB}{105,129,241}

\newcommand{\circlednumberblue}[1]{%
  \textcolor{mycolorblue}{\ding{\numexpr171 + #1\relax}}%
}

\AtBeginDocument{%
  }

\copyrightyear{2025}
\acmYear{2025}
\setcopyright{cc}
\setcctype{by}
\acmConference[ISCA '25]{Proceedings of the 52nd Annual International
Symposium on Computer Architecture}{June 21--25, 2025}{Tokyo, Japan}
\acmBooktitle{Proceedings of the 52nd Annual International Symposium on
Computer Architecture (ISCA '25), June 21--25, 2025, Tokyo,
Japan}\acmDOI{10.1145/3695053.3731035}
\acmISBN{979-8-4007-1261-6/2025/06}

\acmSubmissionID{123-A56-BU3}

\begin{document}

\title{Phi: Leveraging Pattern-based Hierarchical Sparsity for High-Efficiency Spiking Neural Networks}

\author{Chiyue Wei}
\authornote{Both authors contributed equally to this work.}
\affiliation{%
  \institution{Duke University}
  \city{Durham}
  \country{USA}
}
\email{chiyue.wei@duke.edu}

\author{Bowen Duan}
\authornotemark[1] 
\affiliation{%
  \institution{Duke University}
  \city{Durham}
  \country{USA}
}
\email{bowen.duan@duke.edu}

\author{Cong Guo}
\authornote{Cong Guo is the corresponding author of this paper.}
\affiliation{%
  \institution{Duke University}
  \city{Durham}
  \country{USA}
}
\email{cong.guo@duke.edu}

\author{Jingyang Zhang}
\affiliation{%
  \institution{Duke University}
  \city{Durham}
  \country{USA}}
\email{zhjy227@gmail.com}

\author{Qingyue Song}
\affiliation{%
  \institution{Duke University}
  \city{Durham}
  \country{USA}}
\email{qingyue.song@duke.edu}

\author{Hai Li}
\affiliation{%
\institution{Duke University}
\city{Durham}
\country{USA}}
\email{hai.li@duke.edu}

\author{Yiran Chen}
\affiliation{%
\institution{Duke University}
\city{Durham}
\country{USA}}
\email{yiran.chen@duke.edu}

\begin{abstract}
Spiking Neural Networks (SNNs) are gaining attention for their energy efficiency and biological plausibility, utilizing 0-1 activation sparsity through spike-driven computation. While existing SNN accelerators exploit this sparsity to skip zero computations, they often overlook the unique distribution patterns inherent in binary activations. In this work, we observe that particular patterns exist in spike activations, which we can utilize to reduce the substantial computation of SNN models. 
Based on these findings, we propose a novel \textbf{pattern-based hierarchical sparsity} framework, termed \textbf{\textit{Phi}}, to optimize computation.

\textit{Phi} introduces a two-level sparsity hierarchy: Level 1 exhibits vector-wise sparsity by representing activations with pre-defined patterns, allowing for offline pre-computation with weights and significantly reducing most runtime computation.
Level 2 features element-wise sparsity by complementing the Level 1 matrix, using a highly sparse matrix to further reduce computation while maintaining accuracy.
We present an algorithm-hardware co-design approach. Algorithmically, we employ a k-means-based pattern selection method to identify representative patterns and introduce a pattern-aware fine-tuning technique to enhance Level 2 sparsity. Architecturally, we design \textbf{\name}, a dedicated hardware architecture that efficiently processes the two levels of \textit{Phi} sparsity on the fly.
Extensive experiments demonstrate that \name achieves a $3.45\times$ speedup and a $4.93\times$ improvement in energy efficiency compared to state-of-the-art SNN accelerators, showcasing the effectiveness of our framework in optimizing SNN computation.
\end{abstract}



\maketitle

\section{Introduction}

With the advancement of artificial intelligence, Spiking Neural Networks (SNNs)~\cite{ghosh2009spiking,tavanaei2019deep,zhou2022spikformer,yao2024spike,lv2023spikebert,bal2024spikingbert}, have emerged as a promising solution for various applications, including pattern recognition~\cite{zhou2022spikformer,yao2024spike} and natural language processing~\cite{,lv2023spikebert,bal2024spikingbert,xing2024spikelm}, etc. 
SNNs employ spiking neurons inspired by biological brain networks, where neurons react to binary coded spike activations (0 or 1) from previous layers and emit spikes to subsequent layers.
This spike-based information propagation mechanism introduces \textbf{bit sparsity} in SNNs, potentially leading to significant energy efficiency improvements by ignoring the computation corresponding to zero-bit and only performing calculations corresponding to one-bit spikes~\cite{narayanan2020spinalflow,lee2022parallel,liu2022sato}. 

\begin{figure}
    \centering
    \includegraphics[width=1.0\linewidth]{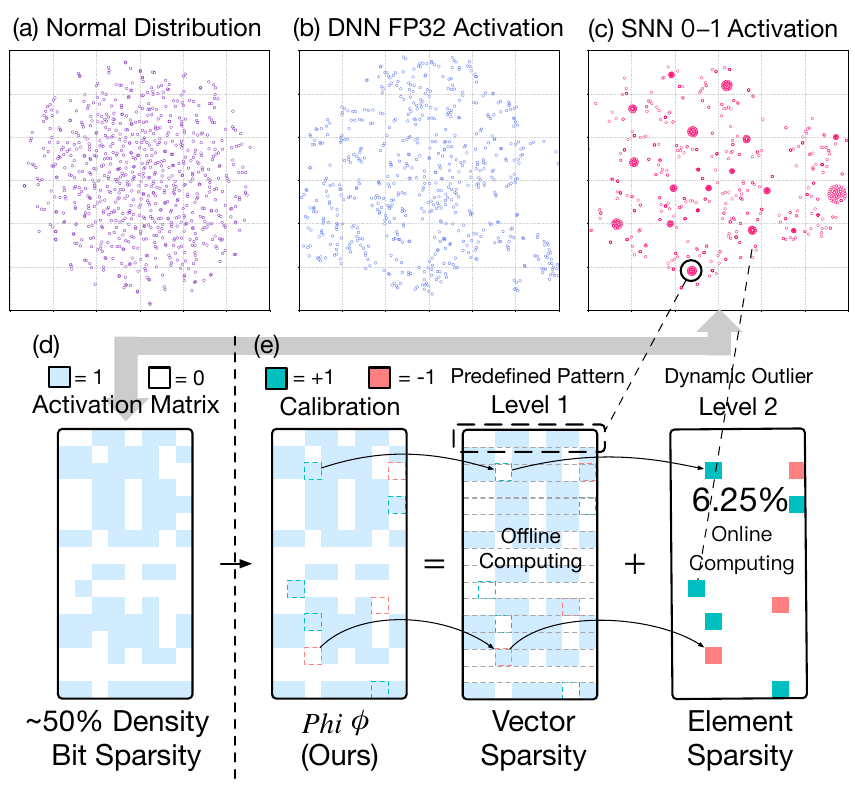}
    \vspace{-10pt}
    \caption{Activation visualization for (a) normal distribution, (b) ViT DNN model, and (c) Spikformer SNN model, and comparison between (d) bit sparsity and (e) ours \name sparsity.}
    \label{fig:intro}
    \vspace{-15pt}
\end{figure}

Extensive studies~\cite{glorot2010understanding,zeiler2014visualizing,he2015delving, lee2017deep} have analyzed the value distribution of Deep Neural Networks (DNNs) activations~\cite{rumelhart1986learning,lecun2015deep}, uncovering useful patterns and developing optimization techniques to enhance efficiency. 
For instance, JPEG-ACT~\cite{evans2020jpeg} observed that activations in Convolutional Neural Network (CNN) models resemble input images, enabling the use of JPEG~\cite{wallace1991jpeg} compression to achieve higher compression ratios. 
Inspired by these advancements in DNN, we observe that spike activations in SNNs exhibit distinct regularities that are often overlooked in prior SNN architectures~\cite{narayanan2020spinalflow,khodamoradi2021s2n2,lee2022parallel,liu2022sato,mao2024stellar}. 
To demonstrate this, we use t-SNE~\cite{van2008visualizing}, a dimensionality reduction technique, to project high-dimensional activations into a lower-dimensional space, comparing the distributions of normal data, DNN activations, and SNN spike activations, as shown in \Fig{fig:intro}. 
Normal distributions appear random and lack structure, while DNN activations are sensitive to specific regions, as previously shown~\cite{glorot2010understanding,zeiler2014visualizing,he2015delving, lee2017deep}. 
In contrast, SNN activations reveal unique structural patterns, presenting new opportunities for architectural and computational optimizations.

Surprisingly, we observe that SNN activations exhibit a much more distinct and regular distribution compared to DNN activations, as illustrated in \Fig{fig:intro}~(c). 
We hypothesize that this phenomenon arises from the binary nature of SNN activations, consisting solely of 0s and 1s, which enforces a more structured distribution. 
Upon closer examination, we identify specific patterns within the SNN activations that can be grouped into well-defined clusters, further highlighting the unique characteristics of SNNs.
Naturally, we can utilize this property to optimize the computation of SNNs.

In this study, we propose a \textbf{pattern-based hierarchical sparsity} framework, termed \textbf{\name} Sparsity, to significantly reduce computations in SNNs. 
As illustrated in \Fig{fig:intro}~(e), \name introduces a two-level sparsity hierarchy: \textbf{ vector sparsity} (Level 1) and \textbf{ element sparsity} (Level 2). 
For Level 1, we propose a calibration method to generate several pre-defined activation patterns from SNN activations. Each row vector in the Level 1 matrix is a pattern.
Since these patterns are formed by activation tensors, their corresponding results can be pre-computed with weights offline, significantly reducing runtime computation for activations matching these patterns.
Thanks to the regular distribution and binary nature of SNN activations, we select only 128 16-bit patterns, enabling a substantial reduction in computational cost, saving 93.64\% of operations over the original SNN. 
However, with only 128 patterns, it is not feasible to cover all activations, necessitating additional mechanisms to address outliers.

Level 2 targets dynamic activation outliers that are not captured by vector sparsity. Unlike traditional bit sparsity, which only includes $1$, our proposed element sparsity incorporates $\{1, -1\}$ for bidirectional correction. 
This bidirectional element sparsity effectively aligns outlier activations with the pre-defined patterns, enabling Level 1 vector sparsity to operate more efficiently. 
By flexibly distributing $\{1, -1\}$ elements, the summation of activations from Level 1 and Level 2 precisely matches the original activation matrix. 
The element sparsity (96.80\%) is significantly higher than the bit sparsity (83.63\%), which ensures less computation.

Designing \name sparsity and its dedicated architecture poses several challenges. 
From the algorithmic perspective, identifying an optimal combination of patterns is critical, as pattern selection significantly impacts the number of outliers and the Level 2 sparsity. 
From the architectural perspective, while patterns are determined offline, the two levels of \name sparsity must be generated dynamically at runtime because activation is dynamic. 
Furthermore, Level 1 sparsity introduces additional memory overhead, while Level 2 sparsity results in load imbalance on parallel architectures due to its unstructured sparsity mode.

To address these challenges, we propose a k-means-based pattern selection method to identify the optimal pattern combination, minimizing online computation. Additionally, we introduce a pattern-aware fine-tuning method that increases the similarity between activations and patterns, thereby enhancing Level 2 sparsity and reducing outliers. 
On the architecture side, we present \name, a dedicated hardware design that efficiently generates and processes \name sparsity on the fly. \name\ incorporates specialized components to handle both sparsity hierarchies effectively, ensuring balanced memory usage and computational load.

We propose an algorithm-hardware co-design, the \name architecture, which fully exploits the advantages of hierarchical sparsity, achieving significant performance gains. Extensive experiments demonstrate that \name\ achieves a $3.45\times$ speedup and a $4.93\times$ improvement in energy efficiency compared to the state-of-the-art (SOTA) SNN accelerator~\cite{mao2024stellar}.
Our contributions are summarized as follows:
\begin{enumerate}
    \item We propose a novel hierarchical sparsity framework, \name, incorporating vector-wise and element-wise sparsity to significantly reduce computation in SNNs.
    \item We develop a k-means-based pattern selection and pattern-aware fine-tuning algorithm to optimize sparsity efficiency while minimizing outliers.
    \item We design \name, a hardware architecture that dynamically generates and processes two levels of \name sparsity, addressing challenges like memory overhead and load imbalance.
    \item Extensive experiments demonstrate that \name\ achieves $3.45 \times$ and $4.93 \times$ runtime performance and energy efficiency improvements compared to SOTA SNN accelerator~\cite{mao2024stellar}.
\end{enumerate}

\section{Background and Motivation}

\subsection{SNNs Background}

\textbf{SNN basics.}
Spiking Neural Networks (SNNs)\cite{ghosh2009spiking,tavanaei2019deep}, 
inspired by biological neural processes, offer higher energy efficiency compared to traditional Deep Neural Networks (DNNs)~\cite{sze2017efficient,shrestha2019review}.

The key distinction lies in SNNs' temporal processing paradigm~\cite{zhu2024TCJA}, where information is encoded through single-bit spikes across multiple time steps, rather than multi-bit activations used in DNNs.
In SNNs, spiking neurons replace traditional activation functions, processing the linear combination of inputs and weights at each time step to update their Membrane Potential (MP)~\cite{izhikevich2003simple} and determine whether to generate a spike. 
Among various neuron models~\cite{yang2024fully,gerstner2014neuronal,pearson2007implementing,izhikevich2003simple,burkitt2006areview}, this work focuses on the most widely adopted Leaky-Integrate-and-Fire (LIF) model~\cite{pearson2007implementing,gerstner2014neuronal}.

\textbf{SNN efficiency.} 
The computational efficiency of SNNs primarily stems from their binary spike representation (1 for spike, 0 for non-spike), termed bit sparsity. 
In this SNN matrix multiplication, zero-valued elements do not contribute to the output, while one-valued elements correspond to simple accumulations of the corresponding weights. 
Existing SNN accelerators~\cite{narayanan2020spinalflow,liu2022sato,lee2022parallel,mao2024stellar} exploit this sparsity by skipping computations for zero elements and performing only accumulation (AC) operations instead of full multiply-and-accumulate (MAC) operations, thereby achieving significant energy savings.

\subsection{SNN accelerators} \label{sec:baselines}
Most existing SNN accelerators focus on efficiently utilizing the bit sparsity of SNNs. 
SpinalFlow~\cite{narayanan2020spinalflow} 
skips the zero elements in the SNN activation by sorting them chronologically by timesteps and then handling them sequentially.
SATO~\cite{liu2022sato} 
integrates input spikes in parallel at each time step and utilizes a binary adder-search tree to generate output spikes without accumulating membrane potentials. 
PTB~\cite{lee2022parallel} also processes bit sparsity in parallel employing a systolic array-based architecture, although it still handles many inactive time steps.
Stellar~\cite{mao2024stellar} relies on Few Spikes neuron~\cite{stockl2021optimized} to improve the sparsity of activations, along with dedicated dataflow to exploit the sparsity.
Prosperity~\cite{wei2025prosperity} proposes product sparsity to improve the bit sparsity by reusing previous inner product results.

While previous works have focused on optimizing bit sparsity processing through either sequential or parallel approaches, they overlook the inherent patterns within sparse activations. Our \name identifies and leverages these structured patterns, opening new avenues for enhancing SNN computational efficiency.
\subsection{Motivation}

\textbf{Uncovering Inherent Patterns in SNN Activations.} 
Existing Spiking Neural Network (SNN) architectures~\cite{narayanan2020spinalflow, khodamoradi2021s2n2, liu2022sato, lee2022parallel, mao2024stellar} 
often overlook the inherent regularity within these activations, a phenomenon well-established in Deep Neural Networks (DNNs)~\cite{ghosh2009spiking,tavanaei2019deep,zhou2022spikformer,yao2024spike,lv2023spikebert,bal2024spikingbert}, where activations consistently carry meaningful information rather than random noise. 
Motivated by this observation, we investigated the structural characteristics of SNN activations. We use t-SNE to analyze their distribution and relationships. 

As illustrated in \Fig{fig:intro}~(a), (b), and (c), we visualize single-layer activations from both DNN and SNN implementations of a Vision Transformer (ViT)~\cite{dosovitskiy2020image} model alongside normally distributed noise as a baseline. 
Each circle represents an activation row. 
The results reveal that DNN activations exhibit certain patterns distinct from random noise. Surprisingly, SNN activations demonstrate even stronger regularity, forming distinct clusters. This clustering phenomenon may emerge from the binary nature of SNN activations, which imposes more structured distributions compared to activations in DNNs.

\textbf{Leveraging Patterns for Computational Efficiency.} In \Fig{fig:intro}~(c), several obvious clusters are observed, indicating that activations within the same cluster share similarities. To capture these similarities more efficiently with a smaller width, we divide the 0-1 activation matrix into partitions. We define the combination of 0s and 1s within a row as a \textbf{Pattern}. \Fig{fig:intro}~(c) illustrates numerous activation rows have identical or similar patterns. This inherent structure in SNN activations presents a compelling opportunity for computational optimization.

While activation values are generated dynamically during inference, making it impossible to predict the specific positions of zeros and ones, the statistical distribution reveals that certain activation patterns consistently emerge, albeit at varying positions. By identifying these patterns beforehand, we can pre-compute their matrix multiplication results with the weight matrix during the offline stage. During online inference, when an activation row matches a pre-identified pattern, we can directly retrieve the pre-computed results, thereby eliminating the need for real-time computation. Based on these observations, we propose \textbf{Pattern-based Hierarchical Sparsity (\name)}, a novel approach specifically tailored for SNN activation patterns.

\subsection{Challenges}
Our proposed \name\ architecture offers a significant opportunity for computation reduction by leveraging hierarchical sparsity. 
However, exploiting this opportunity presents several challenges in both algorithms and hardware design.

\subsubsection{Algorithmic Challenges}
Detecting and selecting a limited number of patterns that match the most activation rows is essential. The key challenges in finding optimal patterns are:

\textbf{Uncertainty of Activations.} 
Activations are highly variable and are generated by different inputs at runtime. However, we need to pre-define and identify the patterns offline, during a calibration stage, since activations are unknown beforehand. This requirement imposes high demands on the algorithm's effectiveness and its ability to generalize from offline calibration to online applications.

\textbf{Complexity of Patterns.} 
Activation rows exhibit thousands of combinations, corresponding to $2^L$ possible patterns for a pattern length of $L$. Therefore, the number of pre-defined patterns is critical to the efficiency of our design. The search space for patterns is $O\left(\binom{2^L}{q}\right)$, where $L$ denotes the pattern length and $q$ denotes the number of patterns we want to select. This complexity renders finding a globally optimal combination of patterns an NP-hard problem. Balancing the memory and computational overhead of storing and calibrating patterns with the model's efficiency is a significant challenge.

\subsubsection{Hardware Challenges}
\textbf{Dynamic Activation Patterns.} 
Although patterns can be pre-determined, activations are dynamically generated during inference. This dynamic nature poses fundamental challenges for hardware in the efficient implementation of \name. Therefore, activation identification and processing must be performed at runtime with high throughput without impeding computational processing.

\textbf{Structured Sparsity.} 
Our \name architecture is built upon sparse computation. Level 1 vector-wise sparsity is pre-defined, and its results are pre-computed before computation, introducing significant memory access challenges for transferring these results to the on-chip computing array. The substantial volume of pre-computed results exceeds the on-chip buffer capacity, necessitating frequent off-chip DRAM accesses and thereby increasing memory bandwidth demands.

\textbf{Unstructured Sparsity.} 
At Level 2 element-wise sparsity, the exceptionally high sparsity ratio introduces numerous zero elements accompanied by irregular distributions of nonzero values. This extreme sparsity negatively impacts computational array utilization and can cause significant workload imbalance in parallel SNN processing architectures.

To address these challenges, we present \name, an innovative algorithm-hardware co-design approach that leverages the existence of activation patterns to achieve ultra-high efficiency in SNN processing.

\section{Pattern-based Hierarchical Sparsity}

\begin{figure}
    \centering
    \includegraphics[width=1.0\linewidth]{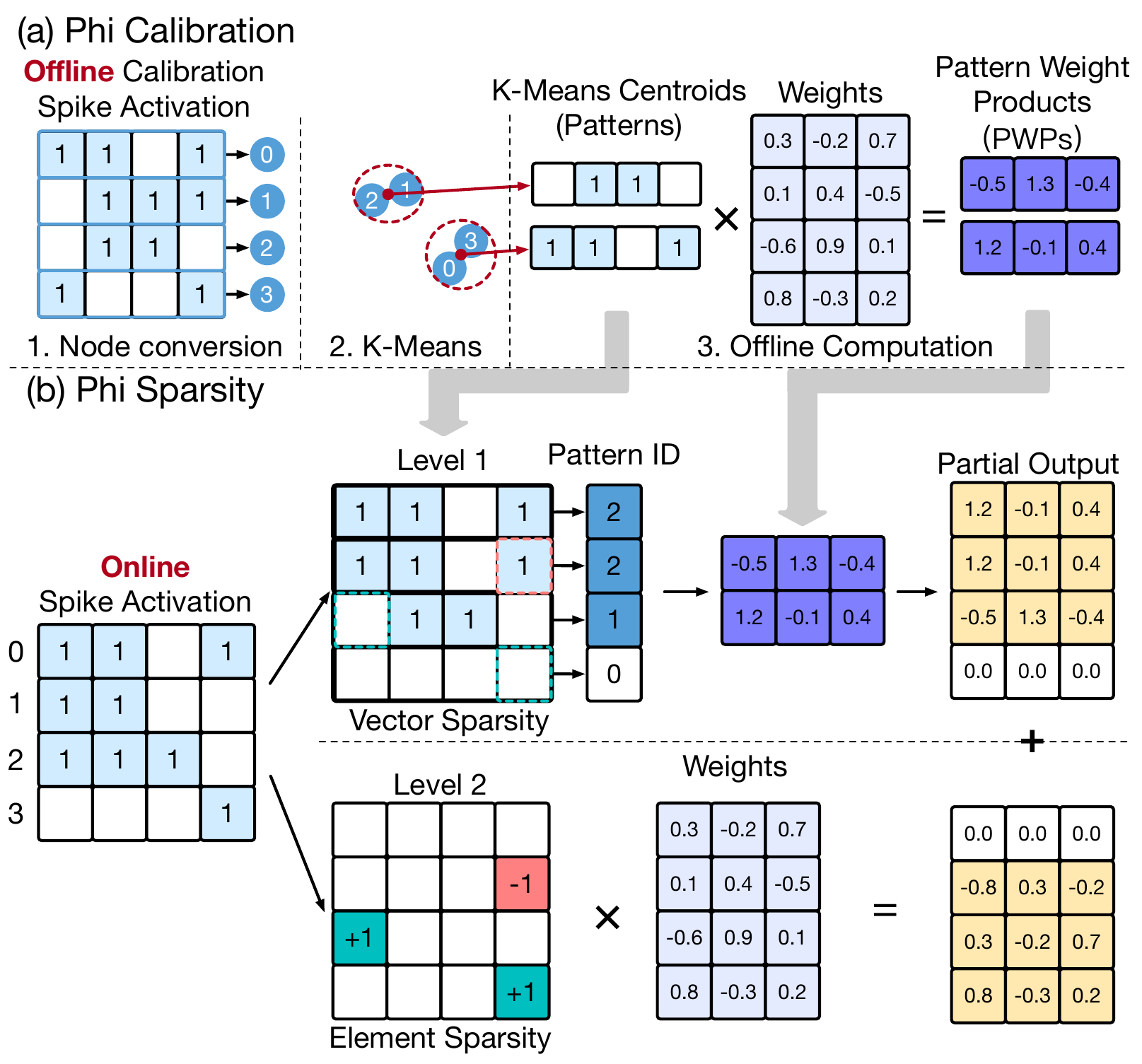}
    \vspace{-10pt}
    \caption{\name Sparsity.}
    \label{fig:phi}
    \vspace{-20pt}
\end{figure}

\subsection{Phi Sparsity definition} \label{sec:phi_definition}
We introduce \textbf{\name} sparsity, a hierarchical approach that leverages patterns in activation to optimize computation. As illustrated in \Fig{fig:phi}, \name sparsity decomposes the activation matrix into two levels of sparse matrices: \textbf{Level 1} comprises pre-defined patterns that are computed offline, yielding structured \textbf{vector-wise} sparsity. In contrast, \textbf{Level 2} captures elements not included in Level 1 and features unstructured \textbf{element-wise} sparsity, which requires runtime processing.

A cornerstone of \name sparsity is the identification of \textbf{patterns} based on the similarity of activation rows. In particular, the similarity between rows is inversely correlated with their length: Longer rows exhibit exponentially more possible configurations, reducing the likelihood of finding patterns that occur frequently. To address this, we divide the activation matrix along the $K$ dimension into smaller partitions of size $k$, identifying patterns within each partition independently.

Our method maximizes the utilization of offline computation to enhance Level 2 sparsity and minimize online computational overhead. The process begins with a \name calibration stage, detailed in \Sec{sec:calibration}, where we define a set of patterns. These patterns will be multiplied with the weight matrix offline to get Pattern-Weight Products (PWPs).

Using this pattern set, we construct two levels of sparse matrices for each activation tile. For each activation row, we identify the best-matching pattern from the set and assign it as the Level 1 sparse row. In straightforward cases, where a row is identical to a pattern (e.g., row 0 matches pattern 1), the pre-computed PWP is directly used as the output row, achieving 100\% Level 2 sparsity with no additional computation.

For cases where a pattern is similar but not identical to an activation row, we employ a bidirectional correction scheme in Level 2 sparsity to handle discrepancies. This scheme addresses two scenarios:

1. \textbf{One to zero mismatch}: When a bit position in the activation is 1 but 0 in the pattern (e.g., \Fig{fig:phi}(b), row 2 = $\underline{1}110$, pattern 1 = $\underline{0}110$), we set a 1 at the corresponding position in the Level 2 matrix as a correction term (e.g., \underline{1}000);

2. \textbf{Zero to one mismatch}: When a bit position in the activation is 0 but 1 in the pattern (e.g., row 1 = $110\underline{0}$, pattern 2 = $110\underline{1}$), we set a -1 at the corresponding position to ensure correctness (e.g.000\underline{-1}).

The Level 2 sparsity is defined as the ratio of matching bits to total bits. For instance, row 1 ($1100$) matching with pattern 2 ($1101$) has a sparsity of 75\% due to 3/4 matching bits. Through the bidirectional correction scheme, we can assign any pattern to an activation row, though the achievable Level 2 sparsity varies. To optimize performance, we assign the pattern that yields the highest Level 2 sparsity. If this sparsity is lower than the row's original bit sparsity (e.g., row 3 in \Fig{fig:phi}(b) has an original bit sparsity of 75\% but the highest achievable Level 2 sparsity of 50\%), the row's original bit sparsity is retained and used in the Level 2 sparse matrix, and no pattern is assigned in Level 1 sparse matrix.

This assignment strategy produces two complementary sparse matrices:

1. The \textbf{Level 1} matrix exhibits vector-wise sparsity, where each row is either a pre-defined pattern or all zeros (indicating no pattern assignment). This structure enables efficient processing. We can transform each row in the Level 1 matrix into the index of the pattern (0 indicates no pattern assigned), and then leverage the pattern index to retrieve the pre-computed PWPs as the output, significantly reducing online computation.

2. The \textbf{Level 2} matrix features \{1,-1\} element-wise sparsity. It maximizes the utility of patterns by incorporating irregularly distributed correction elements to ensure computational correctness while skipping a substantial amount of redundant computation. The Level 2 matrix achieves significantly higher sparsity compared to traditional bit sparsity, allowing for the elimination of substantial redundant computations and greatly enhancing computational efficiency.

Because the activation matrix is tiled (i.e. partitioned) along the $K$ dimension, a tile-wise reduction is performed to aggregate outputs along this dimension. The outputs of Level 1 and Level 2 serve as partial results and are summed to produce the final output.
\subsection{Phi Sparsity Calibration Stage} \label{sec:calibration}
The calibration stage serves as the cornerstone of \name sparsity by calibrating the optimal set of patterns. Pattern selection is crucial as it determines the effectiveness of pattern assignment to activations, whether identical or similar. The similarity between assigned patterns and activations directly impacts Level 2 sparsity. In the calibration stage, we address the challenges of dynamic activation via our distribution analysis and reduce the computational complexity of the calibration stage to linear complexity.

\revise{CQ1}{}\highlight{\textbf{Activation analysis.}
To address the challenges brought by the activation diversity, we investigate the relationship between offline training data and unknown test data distributions. While individual activation matrices may show considerable differences, our analysis reveals that the distribution of activation rows within a partition remains remarkably consistent between training and test datasets, as demonstrated in \revision{\Fig{ft_train_test}}. We empirically determine that a small subset of the training data suffices to represent the distribution of the test data. This finding enables us to perform pattern calibration on a static calibration dataset and effectively apply it during runtime.} \revision{Since activation patterns vary across models, datasets, and layers, we select independent patterns for each model, dataset, layer, and partition to capture unique local distributions.}

\textbf{Efficient Pattern detection.}
Finding the theoretically optimal set of patterns that maximizes Level 2 sparsity for a given activation matrix is an NP-hard problem. However, our investigation of activation distributions reveals heuristics that significantly simplify this process while maintaining high effectiveness.
Based on the clustering behavior observed in \Fig{fig:intro} (c), we identified that the center point of each cluster could naturally serve as the ideal pattern. This approach offers two key advantages:

1. Every activation row belonging to a cluster can be assigned the pattern representing this cluster naturally. Since there are lots of activation rows in each cluster, a majority of rows can be assigned a pattern;

2. By selecting each center point of the cluster as a pattern, we minimize the cumulative distance between the pattern and the activation rows belonging to the pattern, thereby maximizing Level 2 sparsity.

\textbf{Calibration workflow.} Building on the aforementioned analysis, we develop a novel k-means-based clustering algorithm specifically tailored for binary activation patterns, as shown in \Fig{fig:phi} (a). We perform independent k-means clustering for each partition to capture unique local distributions. We first filter out two types of activation rows in the partition since it is meaningless to cluster them and get a pattern for them. 1. All zero rows: all zeros require no computation at all; 2. One-hot row: the Level 2 sparsity of one-hot row cannot be higher than the original bit sparsity unless assigned an identical pattern. 
A one-hot pattern is meaningless since the PWP corresponding to a one-hot pattern is equivalent to a row in the weight matrix.
After the filtering, each row in an activation tile is treated as a k-dimensional binary vector $\in \{0,1\}^k$. To align with our goal of maximizing Level 2 sparsity, we employ the Hamming distance~\cite{jegou2011product} as our distance metric:
$$H(C,D) = \sum^k_{i=0} |C[i] - D[i]|, C,D \in \{0,1\}^{k}$$
The Hamming distance between the cluster center point and data point, representing the number of differing bits between the pattern binary vector and activation row, is directly related to the Level 2 sparsity of the row. Our algorithm design naturally optimizes Level 2 sparsity through k-means iterations. After calculating the center point of the cluster through means, we round the center vectors to 0 or 1 to maintain the binary representation. Our complete k-means-based clustering algorithm is presented in \Alg{alg:kmeans}.

\begin{algorithm}[t]
\caption{K-means-based Pattern Clustering}
\label{alg:kmeans}
\KwIn{Binary dataset $\mathcal{D} = \{x_1, x_2, \dots, x_n\}$, number of clusters $q$}
\KwOut{$q$ representative binary centers}

Initialize $q$ centers randomly from $\mathcal{D}$\;
Filter out one-hot vectors and all-zero vectors from $\mathcal{D}$\;

\For{each iteration}{
  Assign each $x_i \in \mathcal{D}$ to the nearest center using Hamming distance\;
  Update each center as the mean of assigned vectors\;
  Round updated centers to binary values $\{0,1\}$\;
}

\Return Final $q$ binary centers\;
\end{algorithm}

The output centers for each tile (partition) are straightly used as a pattern set and each pattern is assigned a pattern index. The patterns in a set form a $q \times k$ matrix and is multiplied with the weight tile to get the PWPs for usage during runtime. 
By applying our \name sparsity calibration with a tile size of \( k = 16 \) and \( q = 128 \) patterns per tile, we achieve an average density of 3.05\% in the Level 2 sparse matrix. It achieves a significant $5.4 \times$ reduction compared to the original activation matrix, which had a density of 16.37\%.

\subsection{Pattern-aware Fine-tuning}

Our \name sparsity, combined with the novel calibration algorithm, provides a lossless method for optimizing computational efficiency. While this approach already delivers significant gains due to high sparsity, we aim to push the boundaries further by incorporating a lossy method. 

A straightforward approach is to prune the Level 2 sparse matrix by directly setting certain \(1\)s or \(-1\)s to \(0\). However, this naive approach often results in substantial computational inaccuracies, as the weights associated with the pruned values can be large, causing significant deviations in the resulting inner products. To mitigate this issue and ensure minimal degradation in model accuracy, we propose a novel Pattern-aware Fine-tuning (PAFT) technique to enhance Level 2 sparsity. 

The core idea of PAFT is to fine-tune a trained SNN model with an additional regularization term. This regularization term encourages the spike activations to align more closely with the calibrated patterns. Specifically, in addition to targeting high accuracy as usual, the training objective also integrates a penalty that aims to reduce the Hamming distance between the spike activations and the assigned patterns. By progressively aligning activations with the patterns through backpropagation, PAFT improves computational efficiency with minimum compromise of model accuracy.

PAFT is applied after the \name calibration stage of an SNN model. 
To compute the regularization term, we follow the pattern assignment rules outlined in \Sec{sec:phi_definition}. Each row in every partition of the activation matrix is assigned a pattern, and the Hamming distance between the assigned pattern and the corresponding activation is calculated. The regularization term is then defined as the sum of these Hamming distances, which directly correlates to the number of nonzeros in the Level 2 sparse matrix. 

Since our goal is to optimize computation by enhancing Level 2 sparsity, the regularization term is weighted by the \(N\)-dimension of the matrix multiplication at each layer. This weighting ensures the regularization term is directly proportional to the computational cost associated with Level 2 sparsity. The formula for the regularization term is as follows:

\[
R = \sum_{l \in \text{Layers}} N_l \sum_{i=0}^M \sum_{j=0}^{K/k} H(\text{Act}[i, j:j+k], \text{Pattern})
\]

The final loss function incorporates this regularization term with a balancing hyperparameter \(\lambda\):

\[
\text{Loss} = \text{Loss}_{\text{original}} + \lambda\cdot R
\]

This regularization term complements the original loss (e.g., cross-entropy), enabling a trade-off between computational efficiency and model accuracy by tuning \(\lambda\).

Our proposed PAFT is general and can be seamlessly integrated into any SNN model with backpropagation training. Moreover, PAFT is lightweight and efficient, instead of retraining the model from scratch, it requires only a few epochs (e.g., 5) of fine-tuning to achieve $1.26 \times$ of runtime improvements during inference.

\subsection{\revision{Phi Framework Overall Workflow}}
\revise{RB2}{Our \name framework takes a pre-trained SNN model and the training dataset as input. Phi calibration is performed on a small subset of the training dataset to derive patterns specific to the model and dataset. Next, PAFT is applied to the pre-trained model to further align activations with the calibrated patterns.}  

\revision{The calibrated patterns and model weights are then fed into the \name architecture, which generates and processes two levels of sparsity. Notably, PAFT is an optional step that can be skipped if accuracy is a priority. We provide two versions of \name: one with PAFT for improved efficiency and one without PAFT for accuracy-sensitive applications.}
\section{Phi Architecture Design}

\subsection{Overview}

\begin{figure}
    \centering
    \includegraphics[width=1\linewidth]{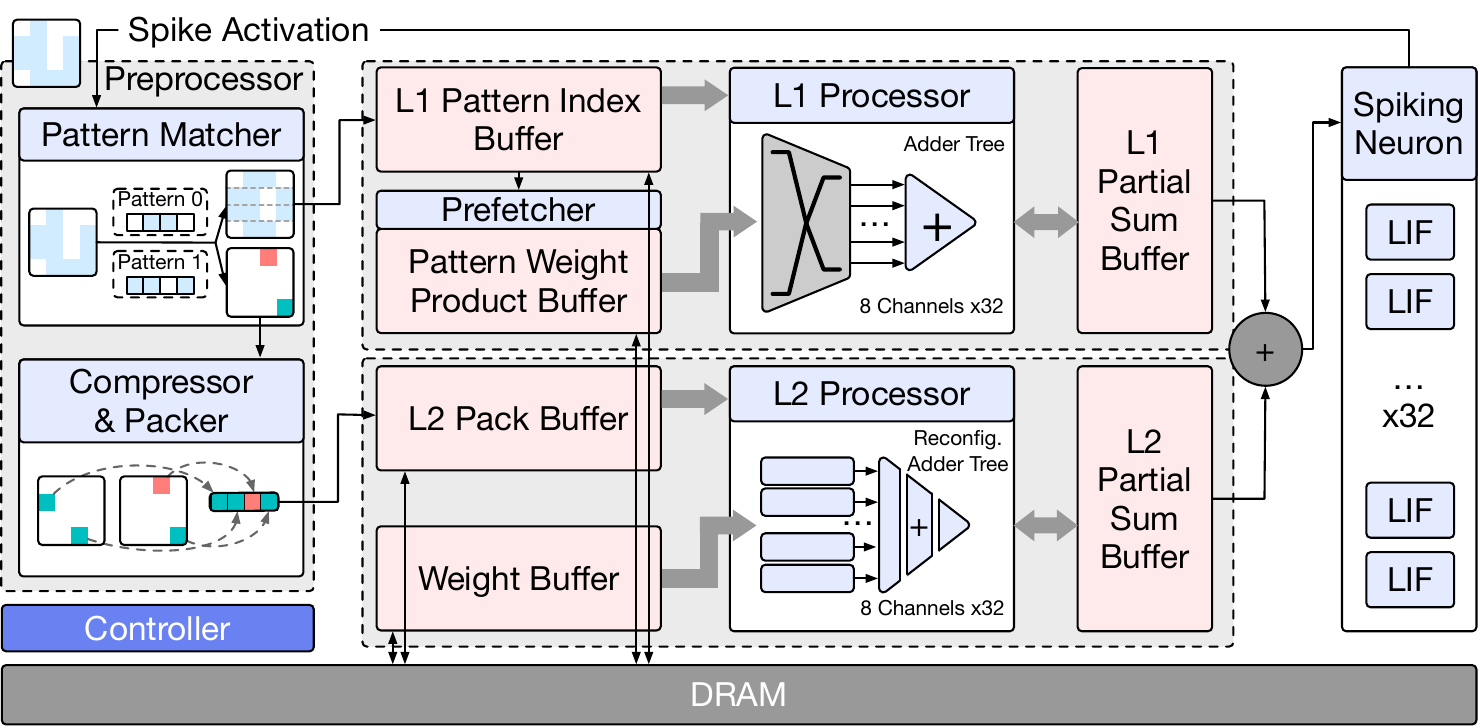}
    \vspace{-10pt}
    \caption{\name architecture overview.}
    \label{fig:overview}
    \vspace{-10pt}
\end{figure}
We present \name, a novel architecture dedicated to our proposed \name sparsity, which efficiently addresses the challenges in exploiting \name sparsity.

Our \name architecture dynamically generates \name sparsity information and efficiently processes two distinct levels of sparsity. The overview of the \name architecture is shown in \Fig{fig:overview}.
The architecture comprises four main components: the Preprocessor, the L1 Processor, the L2 Processor, and a Spiking Neuron Array. The Preprocessor, L1 and L2 processors are optimized for specific aspects of sparse processing.

\textbf{Tiling Strategy.}
We adopt tiling strategy~\cite{chen2014diannao,chen2016eyeriss,gao2019tangram}, a fundamental technique in DNN accelerators that partitions computations to maximize data reuse in on-chip buffers while minimizing expensive off-chip memory accesses. 
We perform tiling across all three dimensions (M, N, K) of the matrix multiplication operation, with a carefully designed execution schedule that prioritizes K-dimension traversal. This K-first ordering enables early reduction operations, which is particularly advantageous for the \name architecture.
The tiling strategy enables each computational round to produce an output tile that flows directly to the Spiking Neuron Array. The resulting spike activations are immediately forwarded to the Preprocessor to process activation for the subsequent layer, enabling a highly efficient overlap of three key operations: matrix multiplication, spiking neuron computations, and preprocessing. This pipelined execution effectively eliminates preprocessing overhead, as these operations are seamlessly integrated into the main computational flow rather than occurring as separate, sequential steps.

\textbf{Preprocessor.}
The Preprocessor transforms spike activations into our two-level \name sparsity representation and optimizes the representation of Level 2 sparsity for efficient processing. 
It employs a pattern matcher that compares predefined patterns with each activation row. For each row, the pattern in Level 1 is determined by identifying the best-matching stored pattern, which simultaneously generates the corresponding Level 2 sparsity information. To help address the challenges of unstructured sparsity of Level 2, including workload imbalance and irregular memory access patterns, the compressor and packer modules consolidate multiple rows of the Level 2 sparse matrix in an optimized scheduling order.

\textbf{L1 Processor.}
The L1 Processor exploits \textbf{structured vector sparsity} in Level 1 through efficient pre-computed Pattern-Wise Products (PWP). It primarily performs PWP retrieval and accumulation based on identified patterns. For rows without matching patterns, the L1 processor implements an efficient skipping mechanism. To solve the challenge of massive off-chip memory access induced by the pre-computed PWPs, we employ a prefetcher to filter out the redundant memory access.

\textbf{L2 Processor.}
The L2 Processor handles \textbf{unstructured element sparsity} using the optimized packed data structure from the Preprocessor, enabling efficient parallel processing through structured memory access patterns and computation scheduling.
While L1 and L2 processors operate independently and concurrently, they require synchronization in each output tile. Our design carefully balances computational loads between processors to minimize inter-processor interference while maximizing overall throughput.

\textbf{Spiking Neuron Array.}
The Spiking Neuron Array processes the aggregated outputs from both processors, implementing the LIF neuron operation to generate spike activations for subsequent layers. This component, while straightforward, is essential for SNN accelerators.

\subsection{Preprocessor}
The Preprocessor has three main parts: Matcher, Compressor, and Packer.

\subsubsection{Phi Pattern Matcher.} 
\begin{figure}
    \centering
    \includegraphics[width=0.9\linewidth]{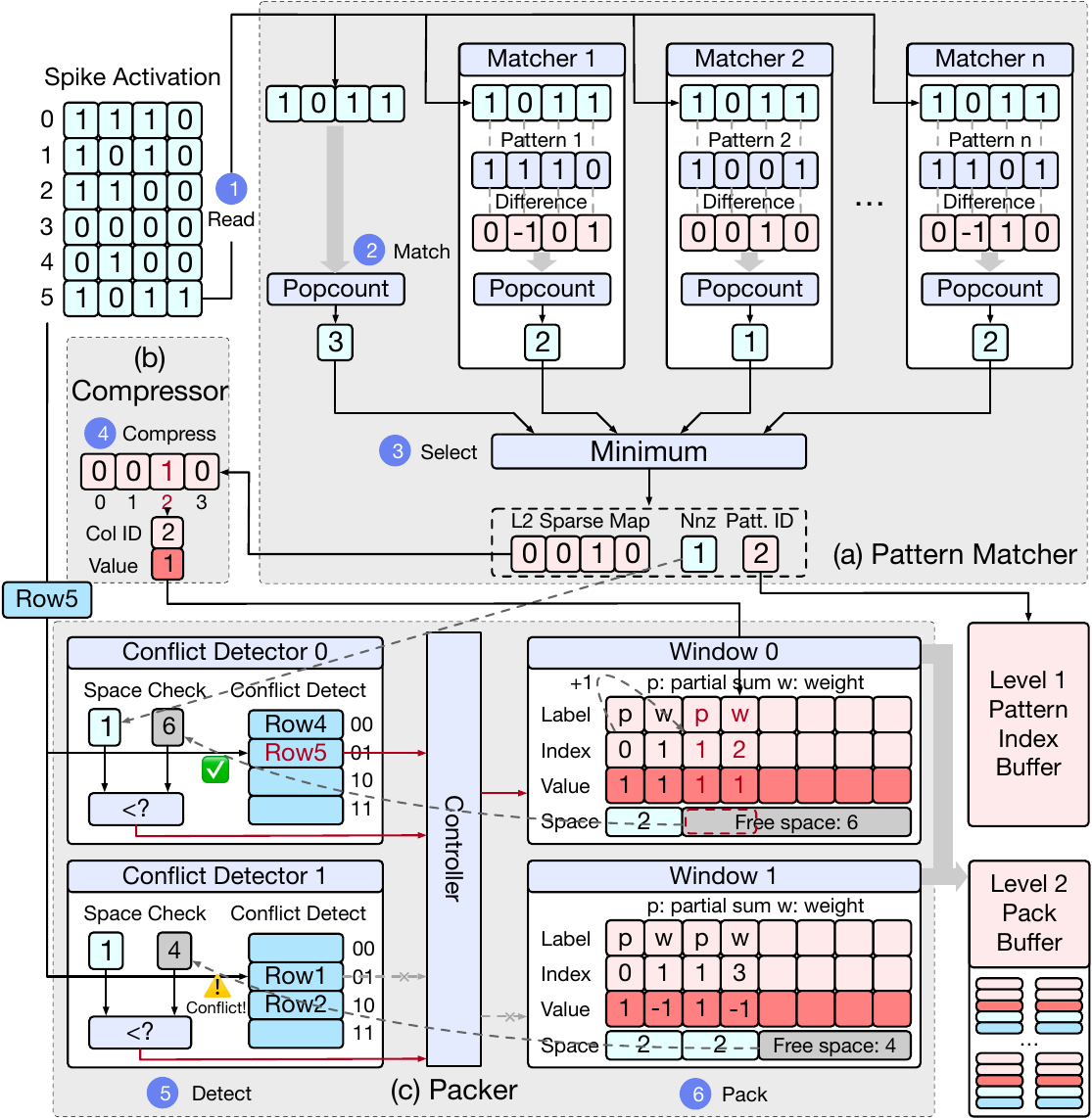}
    \vspace{-10pt}
    \caption{Preprocessor.}
    \label{fig:preprocessor}
    \vspace{-15pt}
\end{figure}

The pattern matcher, a crucial component of the Preprocessor shown in \Fig{fig:preprocessor} (a), implements the \name sparsity mechanism defined in \Sec{sec:phi_definition}. The Matcher processes pre-loaded offline-calibrated L1 patterns for the current tile to generate optimal two levels of sparse representations.

The pattern matching process consists of steps \circlednumberblue{1} through \circlednumberblue{3}. 
In step \circlednumberblue{1}, a spike row from the current activation tile is broadcast to all Matcher units. 
During step \circlednumberblue{2}, each Matcher unit computes the difference between its stored pattern and the input spike activation to generate a candidate L2 sparse row. 
For instance, as shown in \Fig{fig:preprocessor} (a) Matcher 1, the operation $1011 - 1110 = 0,\text{-}1,0,1$ produces a candidate sparse map. 
Popcount operations~\cite{agrawal2019xcel-ram} are then performed on the difference to count nonzero elements in the candidate sparse map and on the original activation to establish a baseline case (representing no pattern assignment). In step \circlednumberblue{3}, the system aggregates Popcount results from all matcher units to identify the minimum count, which corresponds to the highest achievable Level 2 sparsity. This determines both the best-matching pattern and its corresponding Level 2 sparse row.

To optimize hardware efficiency while maintaining functionality, we implement the matcher units in a 1-D systolic array architecture. This design choice offers several key advantages. It maintains a throughput of one row per cycle while reducing hardware overhead compared to parallel implementations. Although the systolic design introduces additional latency, this overhead is effectively hidden by overlapping the operation with L1 and L2 processing.

The pattern-matching process produces two key outputs. First, the matched pattern index is directly forwarded to the Level 1 pattern index buffer for subsequent processing. Second, the generated Level 2 sparse row and its nonzero count undergo further preprocessing stages.

\subsubsection{Level 2 sparse preprocessing.}
The Level 2 element sparsity in our design reaches 96.80\%, which is significantly higher than the original bit sparsity of 83.63\%. While this increased sparsity creates opportunities for operation reduction, it also introduces architectural challenges. To address these challenges, we propose an innovative preprocessing mechanism specifically designed for Level 2 sparsity. Our analysis reveals that the Level 2 sparse matrix for each tiled row is highly sparse, typically containing either zero or 1-2 nonzero elements due to our pattern selection criteria. This high sparsity can lead to hardware underutilization when attempting to process nonzeros within a row in parallel. Therefore, we introduce a row-packing strategy that combines multiple sparse rows into regular-sized packs, enabling efficient utilization of the processor's computation array.

\textbf{Compressor.}
In step \circlednumberblue{4} shown in \Fig{fig:preprocessor} (b), the compressor performs two functions: (1) it filters out all-zero rows generated by the pattern matcher, and (2) it scans the remaining nonzero elements to extract their column indices, creating a compressed representation of the sparse map.

\textbf{Compact data structure.}
We employ a compact data structure for each pack. This structure serves two purposes: it provides regular-sized inputs to the Level 2 processor, enabling parallel processing of nonzeros for improved hardware utilization, and it efficiently handles partial sums from previous partitions. We treat partial sums as special units within the pack since, like nonzeros, they appear irregularly and require summation with weights.

Each of our compact data structures contains 8 units, where each unit contains three fields:
1. A label field indicating whether the unit represents a nonzero element (requiring weight accumulation) or a partial sum from the previous partition;
2. An index field containing either the column index from the compressor or a partial sum index indicating its position among partial sums in the pack;
3. A value field containing either 1 or -1 for nonzeros, or 1 for partial sums (since they are always accumulated).
Additionally, the structure maintains metadata including the number of units occupied per row and the row indices, which are essential for the L2 processor's operation.

\textbf{Packer.}
The packer optimizes row packing density while minimizing buffer access conflicts, as illustrated in \Fig{fig:preprocessor} (c). It employs multiple windows, each holding an incomplete pack and protected by a conflict detector. In step \circlednumberblue{5}, each conflict detector evaluates incoming compressed rows through two stages:
1. Space check: Verifies whether the window's remaining capacity can accommodate the incoming compressed row;
2. Conflict detection: Examines potential bank access conflicts between existing partial sums in the pack and those of the incoming row.

The conflict detection stage prevents bank conflicts during simultaneous read or write of the partial sum. A central controller aggregates the results from all conflict detectors and determines the optimal window for packing each incoming row. In step \circlednumberblue{6}, the incoming row is packed in, with an entry indicating its partial sum.
If no window satisfies the requirements, the most-filled pack is sent to the Level 2 pack buffer and freed for new data. The employment of multiple windows enables flexible pack scheduling, which maximizes pack utilization while preventing buffer access conflicts. Notice that the situation of that one row exceeding the capacity of a pack does not exist due to our optimized Level 2 sparsity.

\subsection{L2 Processor}
\begin{figure}
    \centering
    \includegraphics[width=0.9\linewidth]{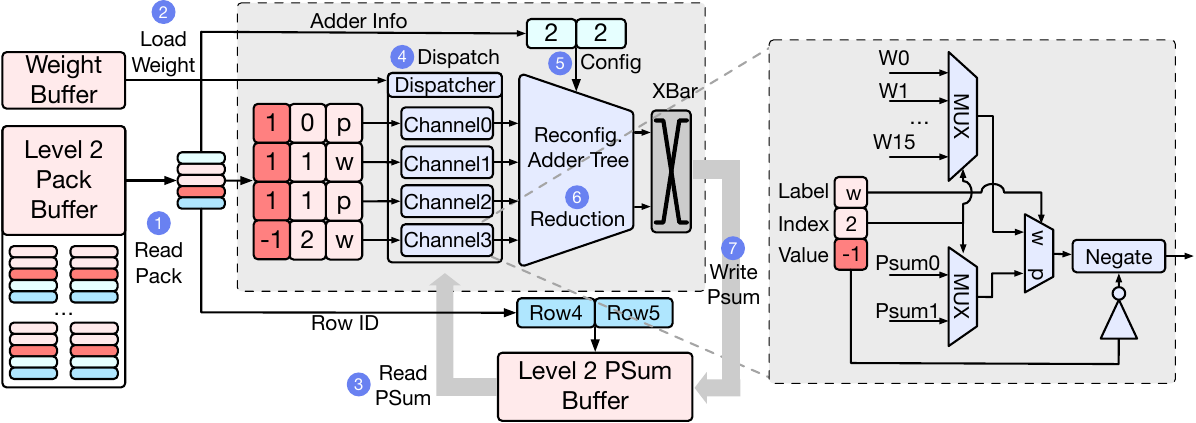}
    \vspace{-10pt}
    \caption{Level 2 Processor}
    \label{fig:l2}
    \vspace{-10pt}
\end{figure}

\begin{figure}
    \centering
    \includegraphics[width=0.9\linewidth]{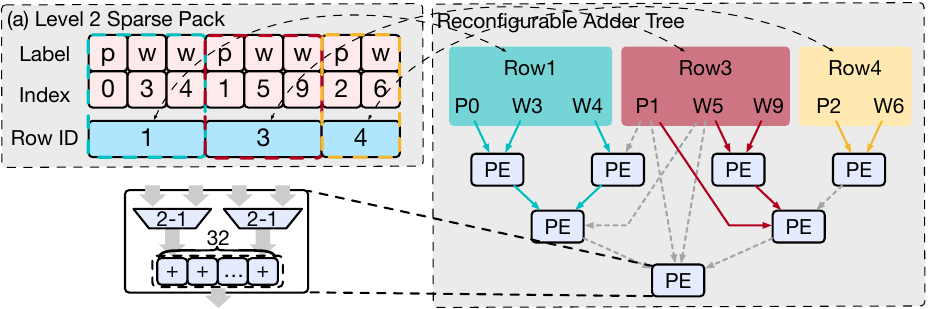}
    \vspace{-10pt}
    \caption{Reconfigurable Adder Tree}
    \label{fig:reconfig}
    \vspace{-10pt}
\end{figure}
Our L2 processor achieves high parallel processing efficiency with optimal hardware utilization, leveraging the organized, compact data structure within each pack. The processing flow, illustrated in \Fig{fig:l2}, consists of seven key steps.
In step \circlednumberblue{1}, the system reads a pack from the L2 pack buffer and decomposes it for distribution to various components. Step \circlednumberblue{2} involves reading a weight tile for the current partition from the weight buffer and forwarding it to the dispatcher. During step \circlednumberblue{3}, the row indices from the pack serve as addresses to retrieve partial sums from the L2 partial sum buffer, which are then sent to the dispatcher. Thanks to our bank-conflict-free packing strategy, partial sum data can be accessed smoothly and efficiently.

Step \circlednumberblue{4} focuses on data preparation in the dispatcher according to the labels, indices, and values from the pack. Each dispatcher channel processes one unit from the pack and feeds one input channel of the adder tree. As depicted in the right section of \Fig{fig:l2}, the label determines whether to use data from the weight buffer or partial sum buffer, while the unit's index selects the appropriate data from multiple partial sums or weights. The selected data undergoes negation if the value is -1, ensuring properly prepared data for dispatch to our reconfigurable adder tree.
The metadata indicating the number of units occupied per row configures our adder tree in step \circlednumberblue{5}. This configuration is crucial since parallel processing of the entire pack requires careful management to ensure correct data summation without cross-row interference. We employ a reconfigurable adder tree structure~\cite{wang2021spatten,qin2020sigma}, shown in \Fig{fig:reconfig}, with 8 input channels. In step \circlednumberblue{6}, this structure enables flexible summation of data belonging to the same output row. The configuration information controls data flow to accommodate varying input quantities and produce the desired number of outputs. For instance, \Fig{fig:reconfig} demonstrates summation of 3, 3, and 2 input channels, respectively. This design introduces only four additional connections compared to a conventional adder tree, resulting in minimal hardware overhead. Each node implements vector SIMD~\cite{hassaballah2008areview} addition for processing row vectors.
Step \circlednumberblue{7} concludes the process by forwarding the adder tree output through a crossbar~\cite{dally2001route} to the appropriate bank in the partial sum buffer.

Our L2 processor achieves high parallelism across all three dimensions (M, N, and K) of matrix multiplication under high sparsity conditions:
M and K dimension parallelism is achieved through the simultaneous processing of units within a pack, handling multiple rows (M dimension) and different units within an activation row (K dimension).
N dimension parallelism is realized through SIMD addition operations on weight and partial sum rows.

This multi-dimensional parallel processing approach, combined with our efficient data organization and reconfigurable architecture, tackles the challenge of workload imbalance and enables high-performance Level 2 sparse matrix multiplication with optimal hardware utilization.

\subsection{L1 Processor}
The L1 processor comprises a 16-to-8 crossbar cascaded with an adder tree for reduction. It reads data from three sources: the Pattern Index Buffer, Pattern Weight Product (PWP) Buffer, and L1 Partial Sum Buffer, computing results that are stored back to the L1 Partial Sum Buffer.

\textbf{L1 Sparsity Management.}
The L1 processor operates on the pattern indices provided by the Pattern Matcher. It reads a pattern index matrix from the L1 Pattern Index Buffer. The whole pattern index matrix records pattern assignments for each tiled row, forming a $M \times K/k$ matrix. 
Our \name sparsity design effectively selects pattern sets and assigns patterns to each row, resulting in a low sparsity pattern index matrix with only $49.34\%$ sparsity on average. Due to this high density, we could employ a straightforward zero-skipping mechanism with little compromise on performance compared with perfect zero-skipping.

The processor examines 16 consecutive elements along the row in the pattern index matrix per cycle, implementing the following logic:
For the situation of a number of nonzeros $\leq 8$: Reads corresponding PWP from respective partitions and forwards them to the adder tree;
For $>8$ nonzeros: Processes the first 8 PWP in the current cycle, with the remaining rows processed in the subsequent cycle.
To enable parallel PWP access across partitions, the PWP buffer is organized into 16 banks, with each partition's PWP stored in a separate bank. A 16-to-8 crossbar connects this buffer to the adder tree, facilitating the sparse skipping mechanism.

\textbf{Memory Traffic Optimization.}
A challenge in L1 processing is the heavy memory traffic induced by PWPs. Employing 128 patterns per K-dimension tile strains the on-chip buffer capacity, making it difficult to store all PWPs for a layer and resulting in additional off-chip memory traffic. Fortunately, our analysis shows that only 27.73\% of precomputed PWPs are utilized within a L1 pattern matrix tile on average, which indicates redundant memory transfers for PWPs.
Moreover, the patterns utilized by the L1 processor are known in advance, thanks to our K-dimension first processing order which generates spike activation and patterns for subsequent layers during current layer computation.
Therefore, we implement a PWP prefetcher, which reads the pattern index and selectively loads necessary PWPs from DRAM, realizing significant off-chip memory reduction.
\section{Evaluation}
In this section, we first detail our evaluation methodology. Next, we explore the design space to justify our configuration. 
We then evaluate \name's performance, energy, and area overhead when applied to SNN models. 
Additionally, we report the improvement of our PAFT technique and illustrate the effect of reducing memory traffic.

\subsection{Methodology}
\label{sec:methodology}
\textbf{Evaluation Models and Datasets.}
We evaluate \name across various spiking CNNs (VGG~\cite{simonyan2014very}, ResNet~\cite{he2016deep}) and spiking transformers (Spikformer~\cite{zhou2022spikformer}, SDT~\cite{yao2024spike}, SpikeBERT~\cite{lv2023spikebert}, SpikingBERT~\cite{bal2024spikingbert}). 
We run these models on tasks including CIFAR10, CIFAR100~\cite{krizhevsky2009learning}, CIFAR10-DVS~\cite{li2017cifar10}, SST-2, SST-5~\cite{socher2013recursive},
and MNLI~\cite{williams2017broad}. 
We obtain the models from open source repositories on GitHub in the paper, train and evaluate them on relevant datasets using PyTorch~\cite{imambi2021pytorch}, and achieve an accuracy close to that reported in the paper. \revise{RB3}{During model training, we adopt the same settings as those described in the paper and the corresponding open source code.
For PAFT after training, we adjust the learning rate (lr) and $\lambda$. A higher learning rate and $\lambda$ will make the pattern in the activations more pronounced, but the accuracy will decrease accordingly. In practice, we search $\text{lr} =1 \times 10^{-5} \sim 1 \times 10^{-3}, \lambda = 0.01 \sim 1$ for each model to achieve the best performance.}
Then, we obtain the model's activations and utilize this information in our experiments.

\textbf{\name Implementation.}
 \revise{RB4}{We developed a simulator based on the methodology of a widely adopted open-source framework~\cite{sharma2018bit,guo2022ant,guo2023olive,wei2025prosperity}. The simulator takes the activations of the model and the calibrated patterns as input, modeling the behavior of each component to estimate the runtime performance of \name.} We implement the \name Preprocessor, the L1 and L2 Processor, and the Spiking Neuron Array using SystemVerilog~\cite{ieee1800_2017}. The components are synthesized with Synopsys Design Compiler in a commercial 28nm process, which provides area and static/dynamic power estimations. We use CACTI 7.0~\cite{muralimanohar2009cacti} to evaluate the area and power of the on-chip buffer in a 28nm process. We use DRAMsim3~\cite{li2020dramsim3} to simulate the off-chip DRAM power. \Tbl{setup} shows the setup of the \name architecture.

\begin{table}
    \centering
    \caption{\name Architecture Setup.}
    \vspace{-8pt}
    \begin{tabular}{l|l}
        \toprule
        \textbf{Parameter} & \textbf{Value} \\ 
        \midrule
        Tile Size& $m=256, k=16, n=32$ \\ \midrule
        On-chip Buffer Size & 4KB Pack; 16KB Weight;\\ &64KB PWP, 28 KB Pattern ID, 128 KB PSum \\ \midrule
        Computation Array & L1 \& L2: 8 channels 32 SIMD adder tree \\ \midrule
        DDR4 DRAM & 8Gb $\times$ 8, 2133R, 4 Channels, 64GB/s \\ 
        \bottomrule
    \end{tabular}
    \label{setup}
    \vspace{0pt}
\end{table}

\textbf{Baselines.}
We compare with \revise{CQ4}{the spiking version of Eyeriss~\cite{chen2016eyeriss} adapted by ~\cite{narayanan2020spinalflow} as a dense baseline} and SNN accelerators
SpinalFlow~\cite{narayanan2020spinalflow},
PTB~\cite{lee2022parallel},
SATO~\cite{liu2022sato},
and Stellar~\cite{mao2024stellar} which we discussed in \Sec{sec:baselines}.

We compare their configuration and performance on VGG-16 with CIFAR100 in \Tbl{tab:baselines}. For a fair comparison, we keep the frequency and technology the same for all baselines and \name.
We define operations (OP) consistently across all baselines and \name, where each OP corresponds to an accumulation for a `1' element in the \textbf{bit sparsity} activation.
We simulate \revision{Spiking Eyeriss}, SpinalFlow, PTB, and SATO in our framework. For Stellar, we rely on the results reported in the paper.

\begin{table}
    \vspace{-10pt}
    \centering
    \caption{Comparison of \name with Baselines. * indicate dependence of specific SNN algorithm.}
    \vspace{-8pt}
    \resizebox{\columnwidth}{!}{
    \begin{tabular}{l|c|c|c|c|c|c}
    \toprule
    & \textbf{Freq.} & \textbf{Tech.} & \textbf{Area} & \textbf{Throughput} & \textbf{Energy Efficiency} & \textbf{Area Efficiency} \\
    & \textbf{(MHz)} & \textbf{(nm)} & \textbf{(mm\textsuperscript{2})} & \textbf{(GOP/s)} & \textbf{(GOP/J)} & \textbf{(GOP/s/mm\textsuperscript{2})} \\
    \midrule
    \textbf{Eyeriss~\cite{chen2016eyeriss}} & 500 & 28 & 1.068 & 9.10 (1.00$\times$) & 5.16 (1.00$\times$) & 8.52 (1.00$\times$) \\
    \midrule
    \textbf{SpinalFlow*~\cite{narayanan2020spinalflow}} & 500 & 28 & 2.09 & 57.23 (6.29$\times$) & 95.77 (18.575$\times$) & 27.38 (3.22$\times$) \\
    \midrule
    \textbf{SATO~\cite{liu2022sato}} & 500 & 28 & 1.13 & 36.01 (3.96$\times$) & 53.22 (10.32$\times$) & 31.86 (3.74$\times$) \\
    \midrule
    \textbf{PTB~\cite{lee2022parallel}} & 500 & 28 & - & 18.12 (1.99$\times$) & 10.65 
 (2.06$\times$) & - \\
    \midrule
    \textbf{Stellar*~\cite{mao2024stellar}} & 500 & 28 & 0.768 & 58.11 (6.39$\times$) & 61.71 (11.96$\times$) & 75.67 (8.89$\times$) \\
    \midrule
    \textbf{\name} & 500 & 28 & \textbf{0.662} & \textbf{242.80} (\textbf{26.70$\times$}) & \textbf{285.81 } (\textbf{55.41$\times$}) & \textbf{366.70} (\textbf{43.06$\times$}) \\
    \bottomrule
    \end{tabular}
    }
    \label{tab:baselines}
    \vspace{-15pt}
\end{table}

\subsection{Design Space Exploration}
\begin{figure}[h]
    \centering
    \includegraphics[width=\linewidth]{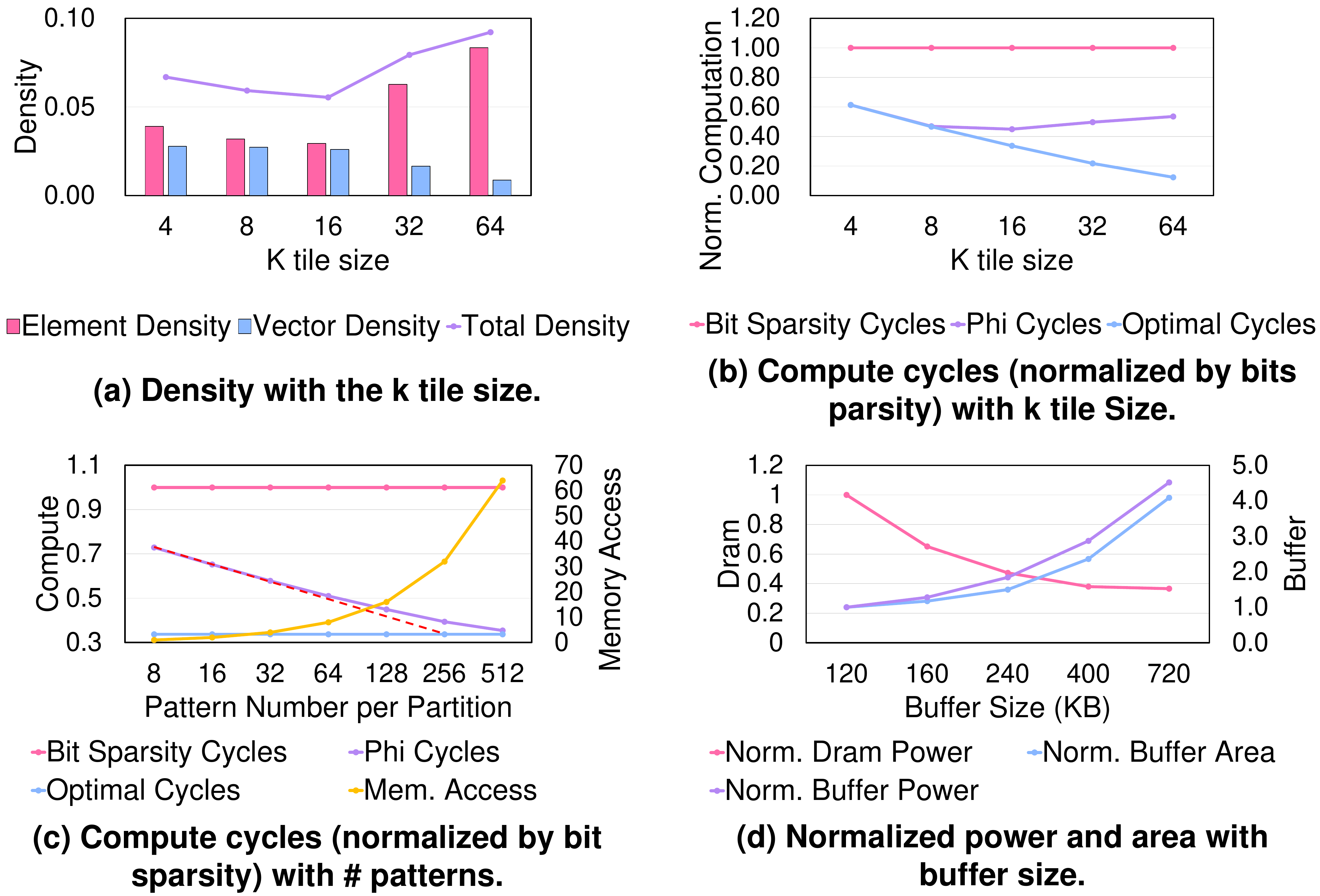}
    \vspace{-10pt}
    \caption{Design space exploration}
    \label{fig:design_exp}
    \vspace{-10pt}
\end{figure}

We explore three key design choices in \name: k-dimension tile (i.e., partition) size, number of patterns, and buffer size to justify our settings in \Tbl{setup}. The experimental results are shown in \Fig{fig:design_exp}.

\begin{figure*}[t]
    \centering
    \includegraphics[width=\linewidth]{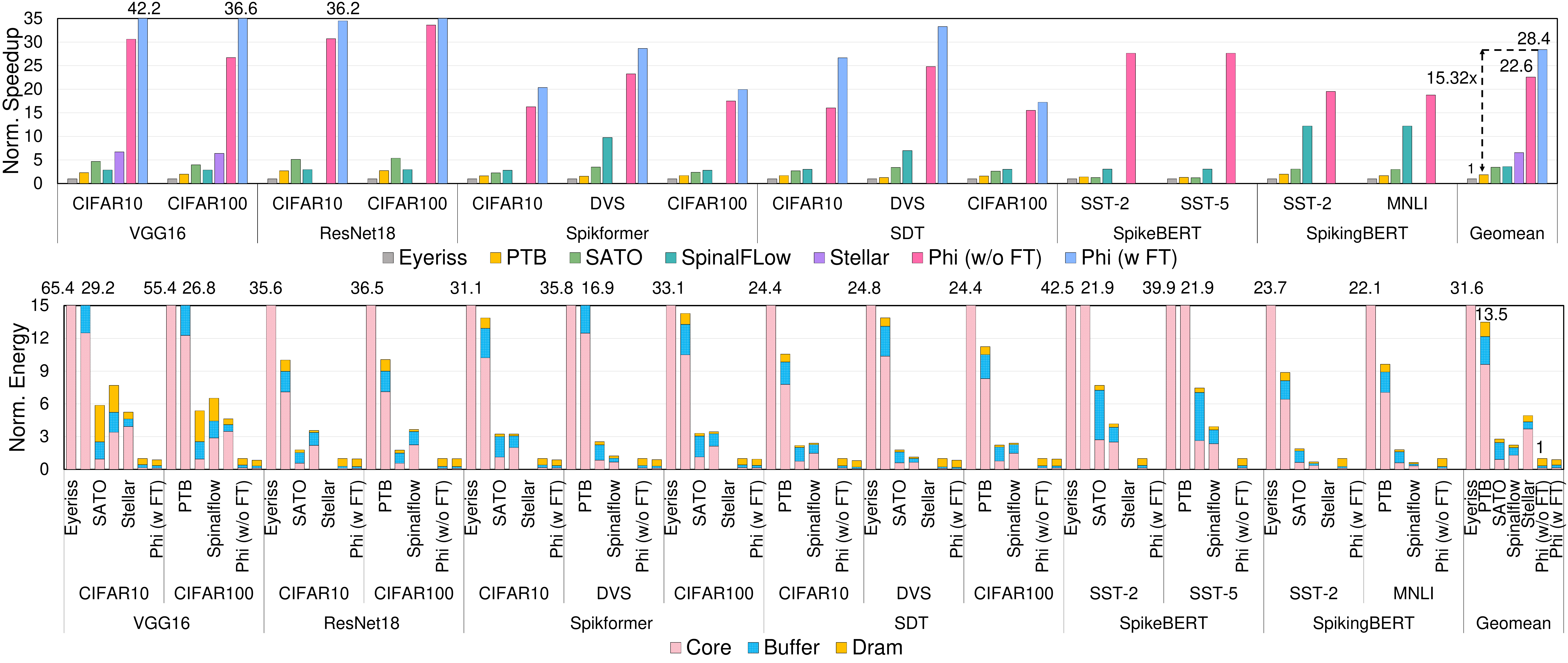}
    \vspace{-15pt}
    \caption{Speedup (normalized by Spiking Eyeriss) and energy (normalized by \name w/o PAFT) of \name and baselines.}
    \label{fig:speed_energy}
    \vspace{-10pt}
\end{figure*}

\subsubsection{\textbf{Tile Size of K}} We first analyze how \textit{K} dimension tile size impacts two hierarchical sparsity and theoretical computation cycles on VGG16 with CIFAR100. We keep the number of patterns per tile at 128. 

\Fig{fig:design_exp}~(a) shows element and vector density with tile size. 
\Fig{fig:design_exp}~(b) shows how \name and optimal cycles change with the tile size. 
Combining \Fig{fig:design_exp}~(a) and \Fig{fig:design_exp}~(b), we can find that the density reflects the amount of computation, and the total computation is the least when k = 16. Meanwhile, we find that k = 16 has the lowest element 
density, which involves the least number of irregular computations and memory accesses. It can reduce the overhead of the L2 Processor part. Finally, we find that the elemental sparsity and vector sparsity are the closest when k = 16, which means that we can allocate the same computational resources to both parts of the processor. This makes our hardware design more balanced and has better hardware utilization and performance when faced with small fluctuations in sparsity across different models and datasets. Therefore, we finally use the k = 16 for both software and hardware design.

\subsubsection{\textbf{Number of Patterns}}
After determining the k tile size, we set the k = 16 and plot the computation cycles and memory accesses versus the number of patterns in \Fig{fig:design_exp}~(c). As the number of patterns increases, the cycles of elements gradually converge towards optimal cycles. When the number of patterns is small, the reduction in computation is nearly linear. However, as the number of patterns increases, the additional memory access required grows rapidly. Therefore, we choose to use 128 patterns to balance computation and memory access.

\subsubsection{\textbf{Buffer Size}}
Finally, since our approach results in additional DRAM access, we investigate the relationship between the size of the on-chip buffer and DRAM power. 
As illustrated in \Fig{fig:design_exp}~(d), increasing the buffer size allows the storage and reuse of more data within the buffer. 
This reduces the amount of access to the DRAM, which in turn reduces power consumption. 
The power of DRAM no longer decreases because the buffer is large enough to load all data at once. 
However, at the same time, a larger buffer size also increases the area and power of the buffer. 
So, we finally set the buffer size = 240 KB to balance the power of DRAM and buffer.

\subsection{Phi Performance and Energy}
\label{sec:performance}
\Fig{fig:speed_energy} and \Tbl{tab:baselines} demonstrate \name's superior efficiency compared to baselines in throughput, area, and power efficiency.

\subsubsection{\textbf{Performance}}
PTB is better than \revision{Spiking Eyeriss} but weaker than other baselines because it does not fully utilize bit sparsity, and there are still zero elements in each time window. SATO leverages bit sparsity, but has some load imbalance issues. Spinalflow demonstrates relatively better performance because it operates under the assumption that each neuron fires at most once during all timesteps. 
However, this assumption negatively impacts the accuracy of SNN models and makes Spinalflow unsuitable for other types of SNN models~\cite{lee2022parallel}. Stellar utilizes software-hardware co-design to achieve and utilize higher sparsity and, therefore, performs best in baselines. \name outperforms SNN ASICs baselines, including PTB, SATO, SpinalFlow, and Stellar, by $12.18\times$, $6.57\times$, $6.29\times$, and $3.45\times$, respectively. This is because of our \name sparsity that decomposes the activation matrix into offline computed Level 1 and highly sparse Level 2, along with our dedicated \name architecture optimized for two levels of sparsity, respectively.

\subsubsection{\textbf{Energy}}
The bottom plot of \Fig{fig:speed_energy} shows the energy of \name and baselines. Overall, \name achieves an energy efficiency of $31.59\times$, $13.48\times$, $2.77\times$, $2.23\times$, and $4.93\times$ when compared to \revision{Spiking Eyeriss}, PTB, SATO, SpinalFlow, and Stellar, respectively. 
Our DRAM energy is not significantly reduced due to additional memory accesses to reduce computation, but despite this, we still achieve a fairly significant energy reduction through our reduced amount of computation.

\subsubsection{\textbf{Area and Power Breakdown}}
\Tbl{tab:area_bd}\revise{RB1}{} presents the area and power breakdown of \name. The on-chip buffer occupies the largest area to minimize memory access. Due to converting computation into memory access, the buffer ranks as the highest power consumer, respectively. Additionally, the area and power associated with the L2 processor are larger than those of the L1, as unstructured sparsity leads to a more complex hardware design.

\begin{table}
    \centering
    \caption{\name Area and Power Breakdown.}
    \vspace{-10pt}
    \label{tab:area_bd}
    \begin{tabular}{lrr}
    \toprule
        \textbf{Components} & \textbf{Area (mm\(^2\))} & \textbf{Power (mW)} \\ 
    \midrule
        Preprocessor & 0.099 & 22.5\\ 
        L1 Processor & 0.074 & 68.2 \\ 
        L2 Processor & 0.027 & 25.6 \\ 
        LIF Neuron & 0.011 & 9.4 \\ 
        Buffer & 0.452 & 220.8 \\ 
    \midrule
        \textbf{Total} & \textbf{0.662} & \textbf{346.6} \\ 
    \bottomrule
    \end{tabular}
    \vspace{-10pt}
\end{table}

\subsection{Pattern-aware Fine-tuning Analysis}
\label{sec:paft_ana}
 \Fig{fig:speed_energy} shows that \name with PAFT achieves $1.26\times$ speedup and $1.1\times$ energy efficiency compared to \name, which we will analyze in the following part. 

\begin{figure}
    \centering
    \subfloat[Training set and Test set without PAFT.]{
        \includegraphics[width=0.3\linewidth]{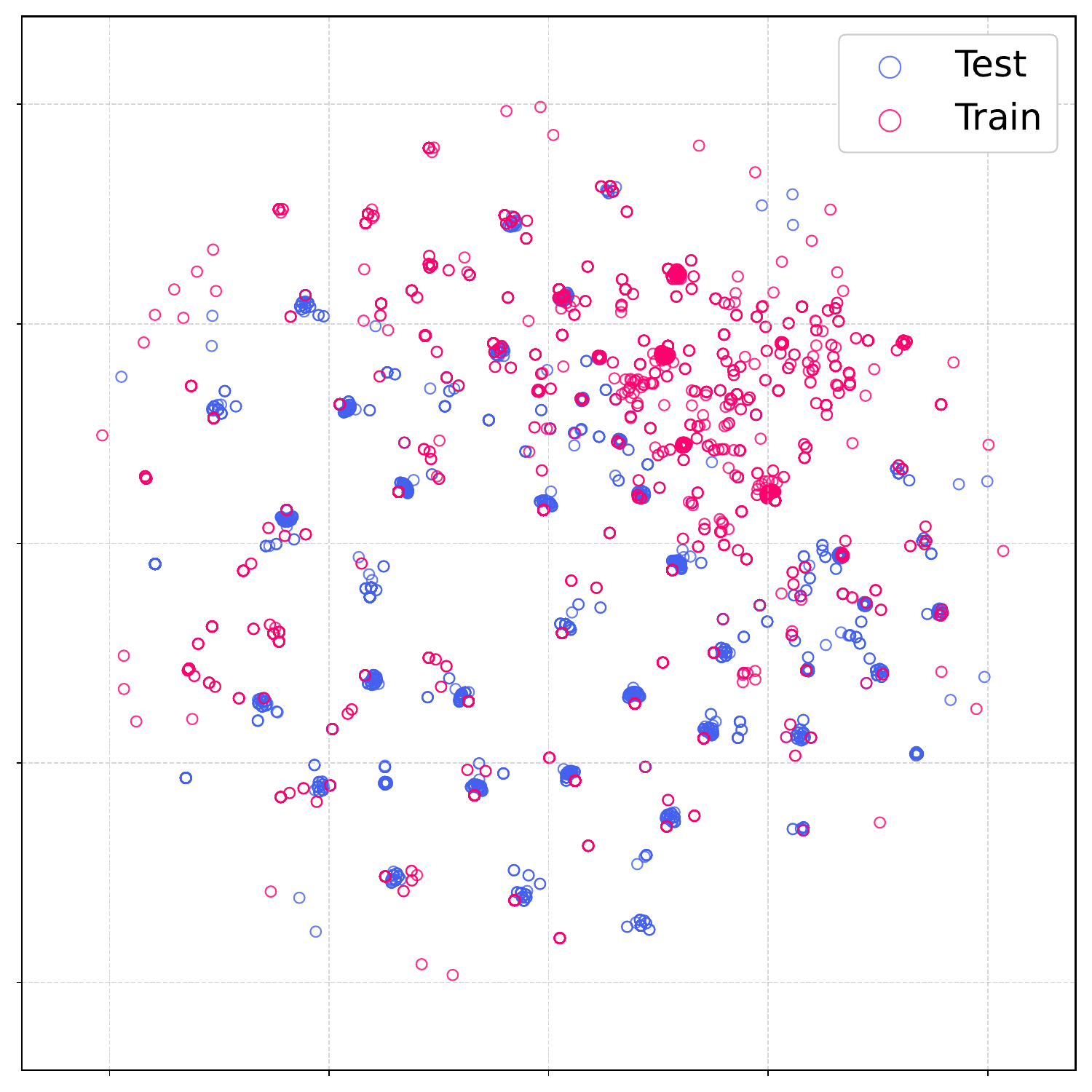}
        \label{ft_train_test}
    }
    \hfill
    \subfloat[Test set without PAFT.]{
        \includegraphics[width=0.3\linewidth]{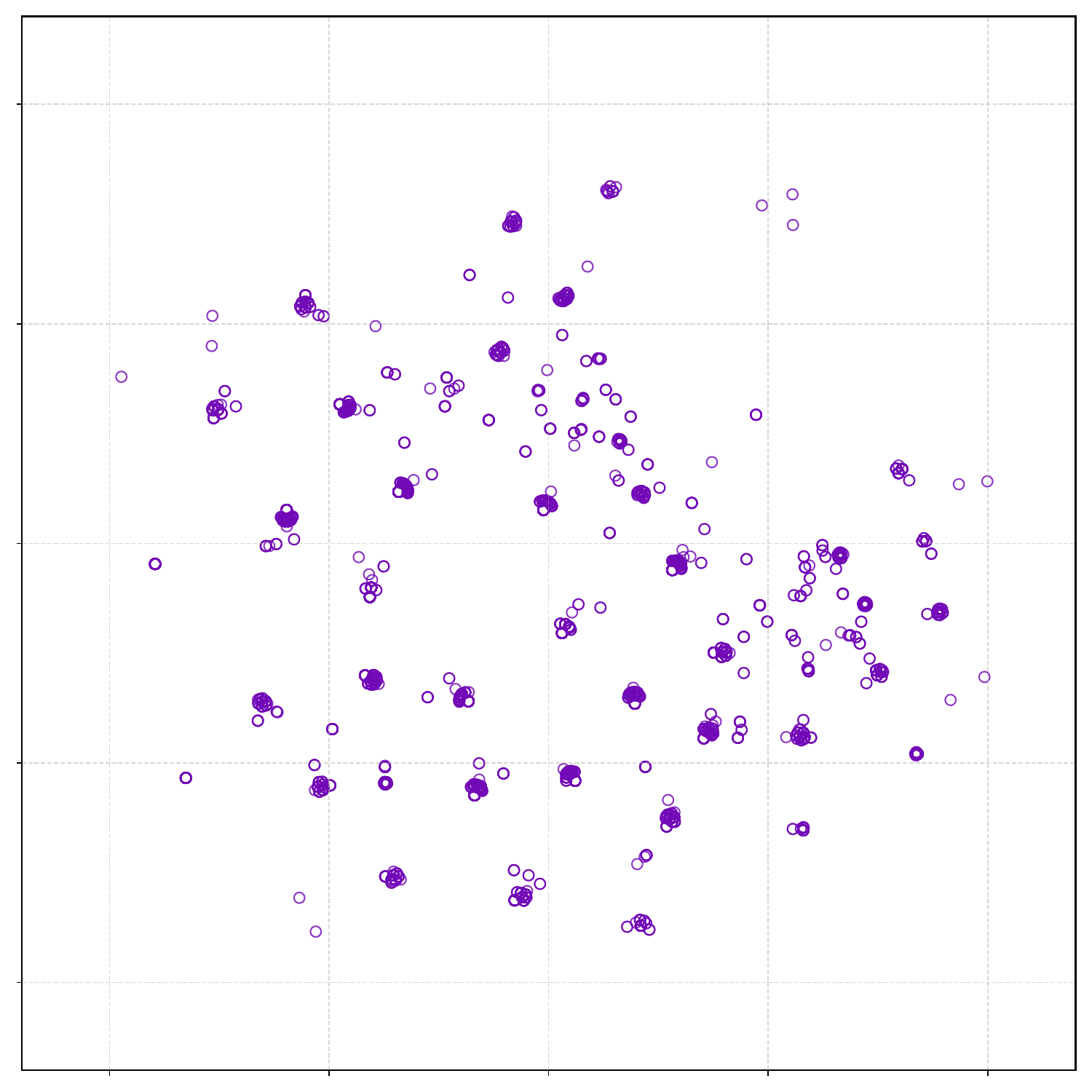}
        \label{ft_test_b}
    }
    \hfill
    \subfloat[Test set with PAFT.]{
        \includegraphics[width=0.3\linewidth]{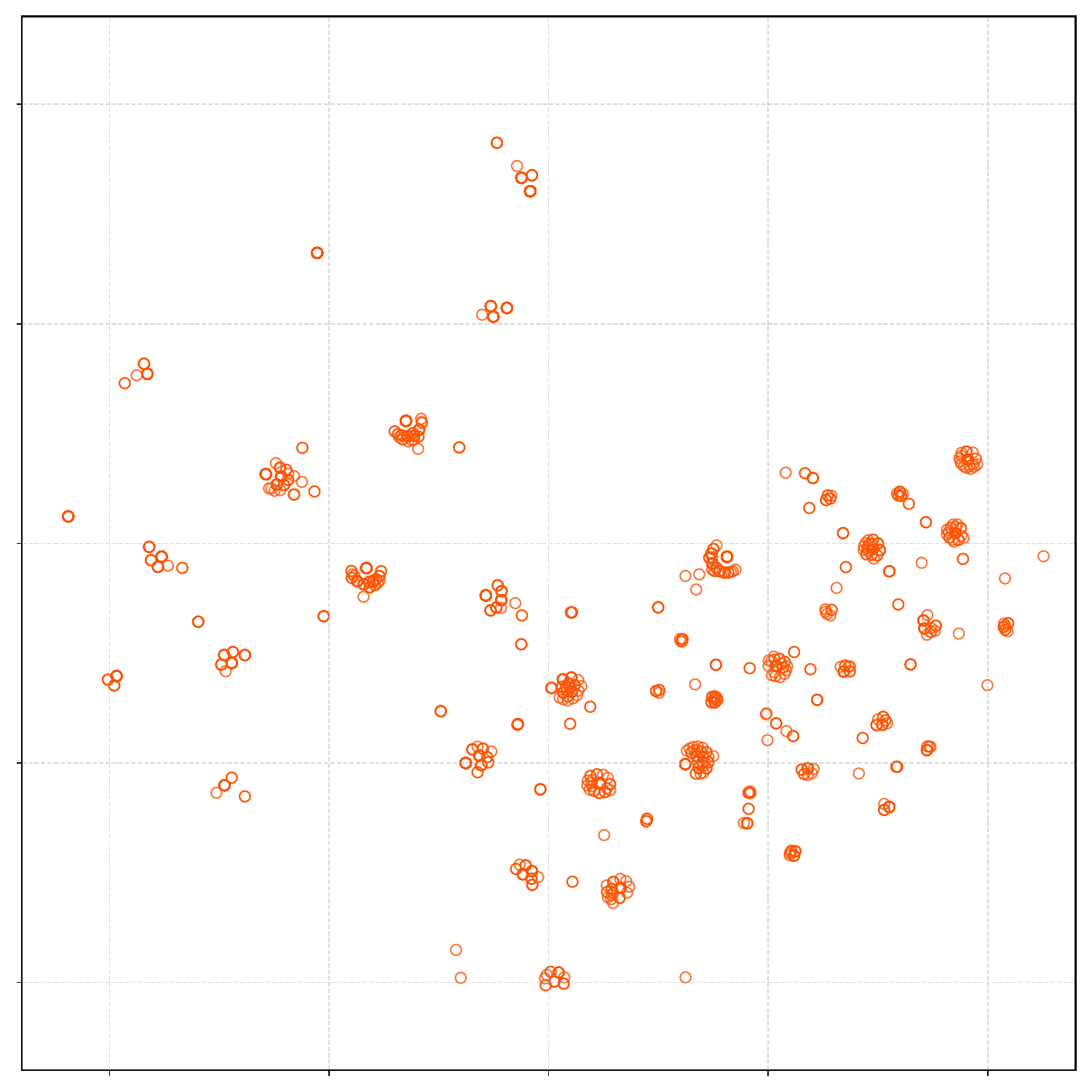}
        \label{ft_test_a}
    }
    \vspace{-5pt}
    \caption{Comparison of t-SNE visualizations for VGG16 on CIFAR-100 without and with PAFT.}
    \label{fig:tsne_comparisons}
    \vspace{-5pt}
\end{figure}

 \begin{figure}
\centering
\includegraphics[width=\linewidth]{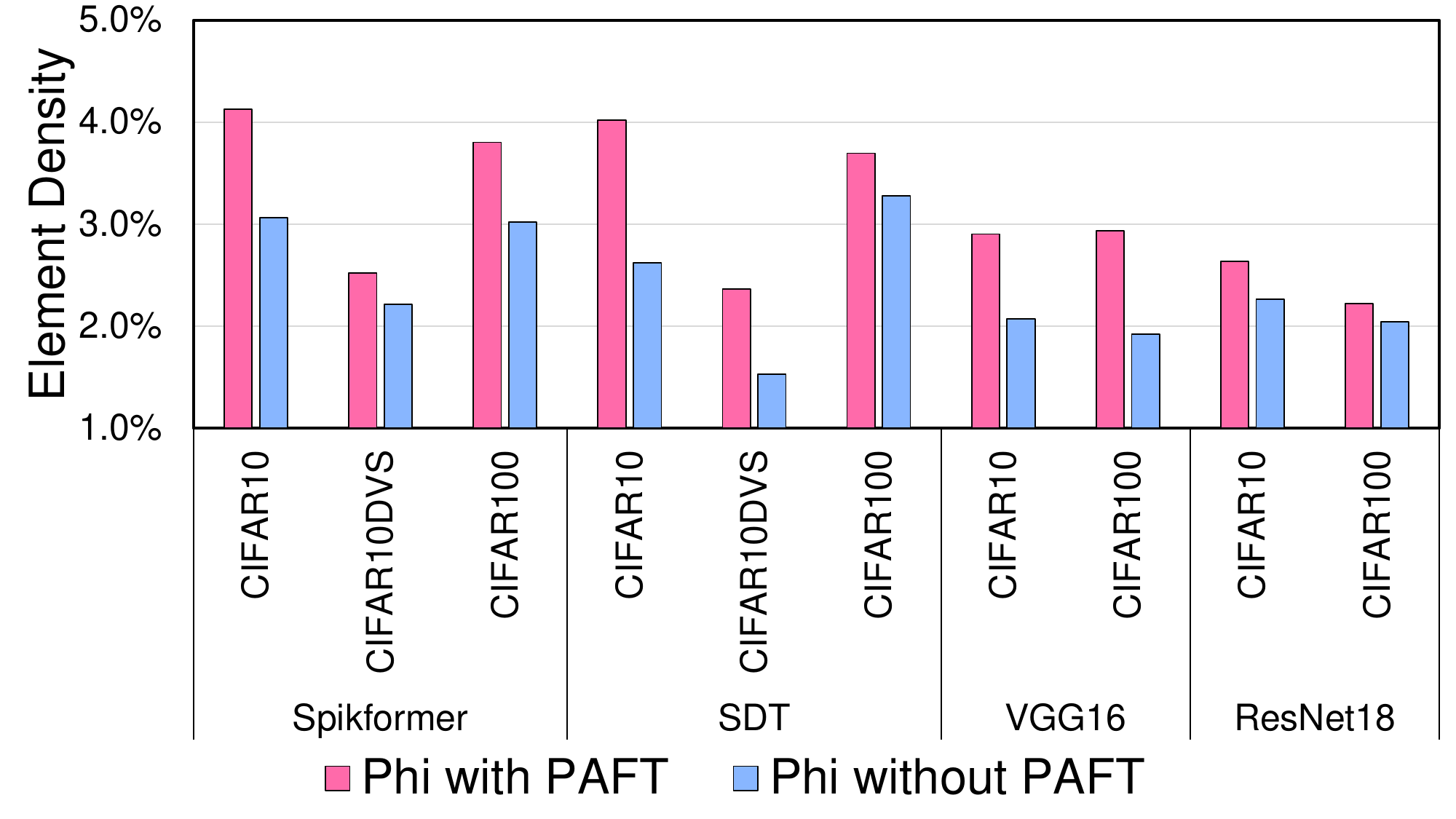}
\vspace{-5pt}
\caption{Element density with and without PAFT.}
\label{fig:ft_sp}
\vspace{-5pt}
\end{figure}

\subsubsection{\textbf{Improvement Analysis}}
\Fig{fig:tsne_comparisons} presents t-SNE visualizations of VGG16's first convolution layer on CIFAR100, demonstrating PAFT's impact. Training set activations effectively encompass test set clusters (\Fig{ft_train_test}), validating our pattern selection approach. The visualizations of the test set with PAFT (\Fig{ft_test_a}) show fewer but denser clusters than activation without PAFT (\Fig{ft_test_b}), indicating that the \name with PAFT better utilizes patterns and reduces the hamming distances. \Fig{fig:ft_sp} reveals decreased element density after PAFT, which significantly affects \name performance given that element sparsity computation is our primary bottleneck.

\begin{figure}
\centering
\includegraphics[width=\linewidth]{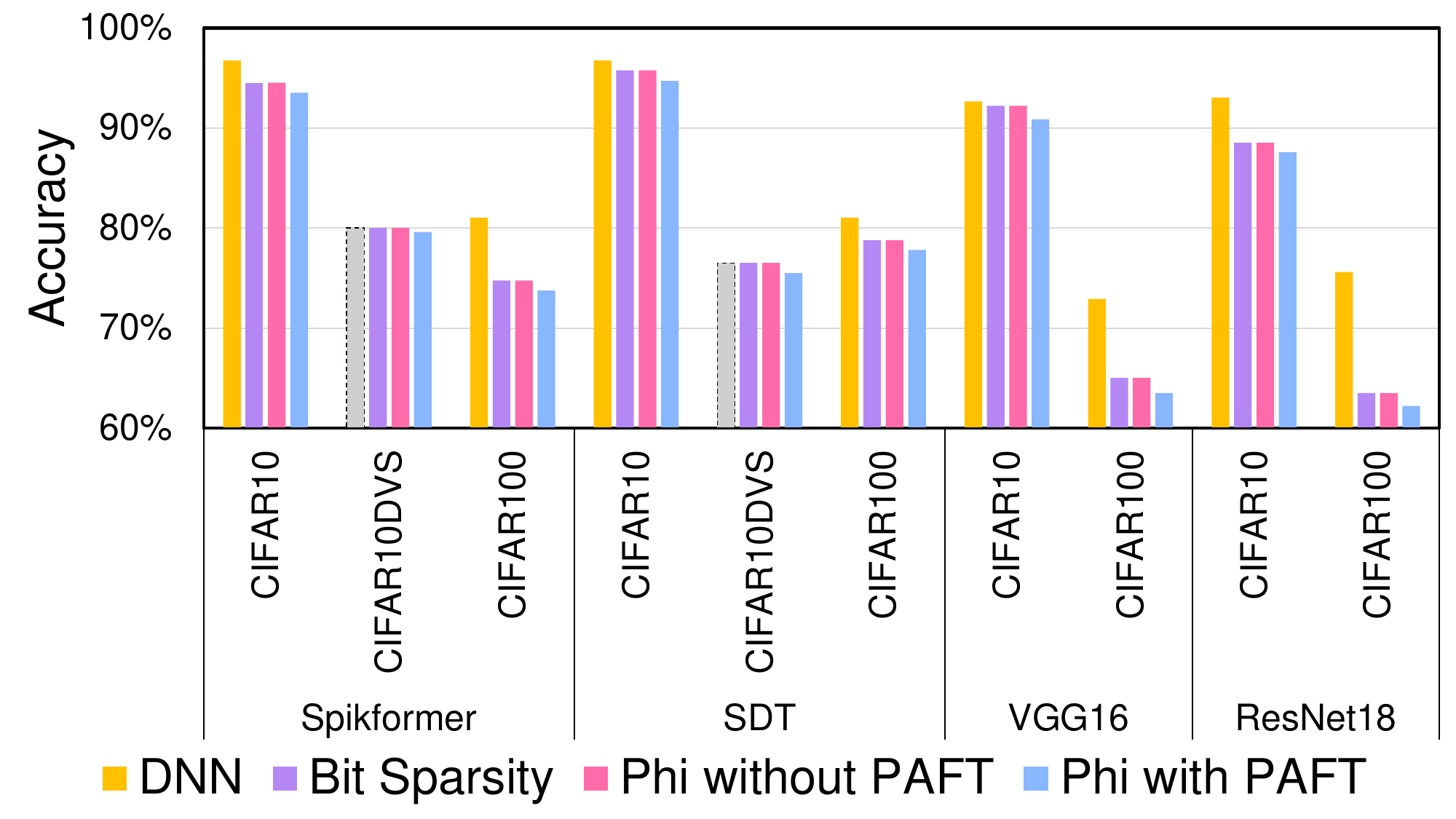}
\vspace{-10pt}
\caption{\revision{PAFT accuracy results.}}
\label{fig:ft_acc}
\vspace{-15pt}
\end{figure}

\subsubsection{\textbf{Accuracy Results}}

\Fig{fig:ft_acc} demonstrates the accuracy of \revision{DNN counter part}, Bit sparsity, \name without PAFT and \name with PAFT across various models and datasets. \revise{RE3}{DNNs outperform SNNs in general tasks. SNNs, however, excel in event-driven tasks (e.g., CIFAR10-DVS) due to their binary spike encoding, where DNNs are not applicable.}
The accuracy of \name without PAFT is equal to that of Bit sparsity, suggesting that \name without PAFT is lossless. Furthermore, there is only a minor decrease in model accuracy after applying PAFT, demonstrating that PAFT successfully trades a small accuracy loss for performance improvement.

\subsection{Memory Traffic Reduction}
\label{sec:traffic_red}
As mentioned in Section 5.4, we reduce memory traffic by compressing activations and pre-fetching pattern data.

\begin{figure}
    \centering
    \includegraphics[width=\linewidth]{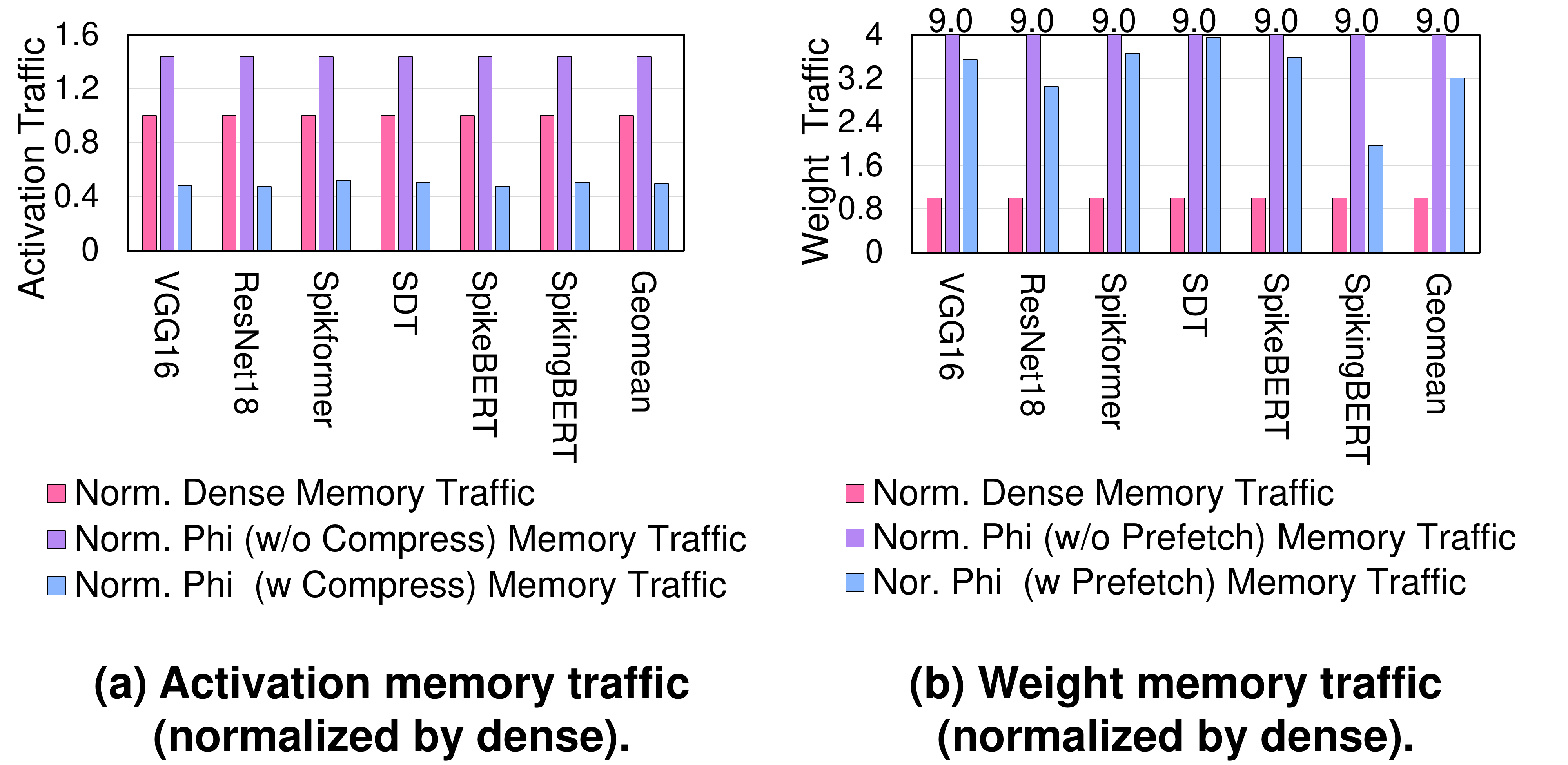}
    \vspace{-10pt}
    \caption{ Effect of reducing memory traffic.}
    \label{fig:mem}
    \vspace{-15pt}
\end{figure}

\subsubsection{\textbf{Compression Rate}}
Our compressor and packer design not only enables better array utilization but also less DRAM traffic. \Fig{fig:mem}~(a) illustrates three types of activation traffic: dense activation traffic (\revision{Spiking Eyeriss}), \name activation traffic without a compact data structure, which requires access to both the element matrix and the corresponding pattern index, and \name activation traffic using a compact data structure. The higher element sparsity provided by \name sparsity allows the compressed format to reduce activation traffic by nearly half, as it does not store zero values.

\subsubsection{\textbf{Prefetcher}}
\revise{RC1}{}\highlight{\Fig{fig:mem}~(b) shows the regular weight traffic (\revision{Spiking Eyeriss}), the \name traffic without prefetching (which requires access to weights and corresponding PWPs), and the \name accesses with prefetching. 
Under our configuration, the weight traffic without prefetching is nine times greater than the normal accesses. 
With prefetching, we reduced the weight accesses to about three times above the regular level, suggesting that some patterns are used frequently, while others are used rarely.}

\subsection{\revision{Phi Generalizability Analysis}}
\label{sec:phi_general}

\textbf{Phi in SNNs.} To evaluate the generalizability of \name, we apply its optimal configuration (partition size $k=16$, 128 patterns per partition) across various SNN models and datasets. As summarized in \Tbl{tab:phi_sparsity}\revise{RA1}{}, \name achieves high L1 density, closely matching the original bit density, and low L2 density. This indicates that predefined patterns effectively represent most activations (computed offline), leaving minimal mismatches (L2 matrix) for online processing. Consequently, \name achieves a theoretical $4.5\times$ speedup over bit sparsity and $38\times$ over dense.

These findings confirm that activation patterns are inherent in diverse SNN models and datasets and \name effectively identifies them.

\textbf{Phi in Random Matrices.} To further explore the universal presence of patterns, we apply \name sparsity to randomly distributed binary matrices with varying densities. The results demonstrate that \name-calibrated patterns consistently lead to low L2 density and a $2.7\times$ theoretical speedup over bit sparsity. This confirms that patterns naturally emerge in binary matrices. However, the theoretical speedup observed in SNN activations is higher than in randomly distributed matrices. This is because SNN activations exhibit a more structured distribution compared to purely random data, making patterns more distinct and allowing \name to achieve better performance.

\begin{table}[t]
\vspace{-5pt}
    \centering
    \caption{\revision{\name sparsity breakdown analysis}}
    \label{tab:phi_sparsity}
    \resizebox{0.95\columnwidth}{!}{
    \begin{tabular}{l|c|c|c|c|c|c|c}
    \toprule
        \multirow{2}{*}{Model} & \multirow{2}{*}{Dataset} & Bit & L1  & L2:+1 & L2:-1 & Theo. Sp. & Theo. Sp.\\ 
        & & Density&Density & Density& Density& Over B. & Over D.\\\midrule \midrule
        \multirow{2}{*}{VGG16} & CIFAR10 & 8.7\% & 7.5\% & 1.4\% & 0.1\% & $5.8\times$  & $66.5\times$  \\ 
        ~ & CIFAR100 & 10.6\% & 9.1\% & 1.6\% & 0.2\% & $5.8\times$  & $54.6\times$ \\ \midrule
         \multirow{2}{*}{ResNet18} & CIFAR10 & 7.4\% & 5.8\% & 1.8\% & 0.2\% & $3.7\times$  & $49.6\times$  \\
        ~ & CIFAR100 & 7.0\% & 5.7\% & 1.6\% & 0.3\% & $3.7\times$  & $52.8\times$  \\ \midrule
        \multirow{2}{*}{SpikingBERT} & SST-2 & 20.3\% & 18.0\% & 3.2\% & 0.8\% & $5.0\times$  & $24.8\times$  \\ 
        ~ & MNLI & 21.0\% & 18.7\% & 3.2\% & 1.0\% & $5.0\times$  & $23.8\times$  \\ \midrule
        \multirow{2}{*}{Spikformer} & DVS & 11.9\% & 10.1\% & 2.2\% & 0.3\% & $4.8\times$  & $39.9\times$  \\ 
        ~ & CIFAR100 & 14.2\% & 11.6\% & 3.3\% & 0.7\% & $3.5\times$  & $24.6\times$  \\ \midrule
        \multirow{2}{*}{SDT} & DVS & 11.2\% & 9.6\% & 1.7\% & 0.1\% & $6.1\times$  & $54.6\times$  \\ 
        ~ & CIFAR100 & 15.2\% & 11.8\% & 4.1\% & 0.7\% & 3.2$\times$  & $20.9\times$ \\ \midrule \midrule
        \multirow{4}{*}{Random} & 5\% & 5.0\% & 2.4\% & 2.6\% & 0.0\% & $2.0\times$ & $39.2\times$ \\
        ~ & 10\% & 10.0\% & 6.6\% & 3.4\% & 0.0\% & $2.9\times$ & $29.6\times$ \\ 
        ~ & 20\% & 19.9\% & 13.9\% & 6.4\% & 0.4\% & $2.9\times$ & $14.8\times$ \\ 
        ~ & 50\% & 50.0\% & 49.8\% & 7.9\% & 7.7\% & $3.2\times$ & $6.4\times$ \\ \midrule
        \bottomrule
    \end{tabular}
    }
    \vspace{-5pt}
\end{table}

\section{\revision{Discussion}}

\subsection{\revision{Benefit and Cost of \name Preprocessing}}
\label{sec:tradeoff}
\revise{CQ3}{}\revision{Our method achieves high L2 sparsity with the support of the \name Preprocessor. To justify our design, we analyze the preprocessing overhead and the corresponding computation savings.  

Although preprocessing, particularly pattern matching, is computationally intensive, requiring comparisons with 128 patterns per operation, it is performed only on activations, resulting in a relatively small number of operations per layer. In contrast, preprocessing significantly reduces accumulation operations in the L1 and L2 Processors, which involve both activations and weights, leading to substantial computational savings.

After averaging experimental data across all SNN models, we find that the energy savings from reduced accumulation operations is $75.5\times$ the energy cost of preprocessing. It demonstrates that \name achieves a highly favorable tradeoff, where the preprocessing cost is justified by a substantial reduction in computational overhead through enhanced sparsity and fewer accumulation operations.
}

\subsection{\revision{Relationship with Sparsity and Quantization in DNNs}}
\label{sec:dnn}

\revise{RA2}{}\revision{

\textbf{Sparsity.} Extensive research has explored the use of sparsity~\cite{qin2020sigma,hua2019boosting,wang2021dual,guo2020accelerating,guo2024accelerating,guo2025survey} in DNNs to improve computational efficiency. While \name draws inspiration from these approaches, it uniquely leverages the binary nature of SNNs to achieve greater performance gains.
For example, CGNet~\cite{hua2019boosting} prunes unimportant channels via structured sparsity, and SIGMA~\cite{qin2020sigma} skips unstructured zero activations.  However, both approaches target zero elements only. In contrast, \name takes advantage of pattern-based sparsity to skip significant computation related to nonzeros in activations, thanks to the unique binary features of SNNs.
}

\textbf{Quantization.} Recent advances in bit-slicing~\cite{chen2024bbs,shi2024bitwave,lu2021distilling,albericio2017bit} for DNN quantization~\cite{lin2024awq,ashkboos2024quarot,guo2022ant,guo2023olive,hu2025m} highlight the potential for generalizing \name beyond SNNs. Bit-slicing decomposes multi-bit integer matrix into a set of binary matrices, enabling the exploitation of bit-level sparsity. Several works leverage this sparsity: BBS~\cite{chen2024bbs} exploits both 1-bit and 0-bit sparsity using symmetry-aware pruning, while Transitive Array~\cite{guo2025transitive} reuses previous results in the binary matrix to reduce redundant computation and enhance bit-level sparsity.
\name, which deals with binary matrices, may be extended to bit-sliced DNNs, offering further opportunities to harness bit-level sparsity.

\section{Conclusion}
This paper presents \name, a novel hierarchical sparsity framework for SNNs that leverages structural patterns in SNN activations to generate two levels of sparsity: vector-wise and element-wise. Our algorithm-hardware co-design approach addresses key challenges through k-means-based pattern selection and pattern-aware fine-tuning, while the dedicated hardware architecture efficiently generates and processes both levels of sparsity on the fly. Experimental results demonstrate that the proposed system achieves a 3.45$\times$ speedup and 4.93$\times$ improvement in energy efficiency compared to state-of-the-art SNN accelerators.
Our work demonstrates the significant potential of pattern-based hierarchical sparsity in advancing the practical deployment of SNNs. Furthermore, this research opens new directions for investigating the inherent activation patterns in SNNs and their potential applications across other neural networks.

\begin{acks}
This work was supported in part by NSF-2112562, NSF-2328805, and ARO W911NF-23-2-0224. The authors sincerely thank the anonymous reviewers for their constructive feedback and valuable suggestions that greatly improved the quality of this work. We also extend our gratitude to Junyao Zhang and Jonathan Ku for their insightful discussions and technical support.
\end{acks}

\bibliographystyle{ACM-Reference-Format}
\bibliography{refs}

\end{document}